\documentstyle[prl,aps,amssymb,graphicx]{revtex}
\newcommand{\beq}{\begin{equation}}
\newcommand{\eneq}{\end{equation}}
\newcommand{\vl}{\vec{\lambda}}
\newcommand{\met}{\frac{1}{2}}
\newcommand{\ro}{ \rho _{\Omega}(s')}
\newcommand{\rot}{ \rho _{\tilde{\Omega}}(s')}
\input{epsf}

\begin{document}

\tolerance 10000

\twocolumn[\hsize\textwidth\columnwidth\hsize\csname %
@twocolumnfalse\endcsname

\draft

\title{Adiabatic Control of the Electron Phase in a Quantum Dot}

\author {D. Giuliano$^{1,2,5}$, P. Sodano$^{3,4}$ and A. Tagliacozzo$^{1,2}$}

\address{$^1$ I.N.F.M. - {\sl  ``Coherentia''}   and Unita' di  Napoli\\
$^2$ Dipartimento di Scienze Fisiche, Universit\`a di Napoli
         ``Federico II'', Monte S.Angelo - via Cintia, I-80126 Napoli, Italy\\
        $^3$ Dipartimento di Fisica, Universit\`a di Perugia, via Pascoli 
            0163, Perugia, Italy\\
        $^4$ I.N.F.N., Sezione di Perugia\\
        $^5$ Dipartimento di Fisica, Universit\`a della Calabria, and Istituto 
	Nazionale di Fisica Nucleare, Gruppo collegato di Cosenza, I-87036
	Arcavacata di Rende, Cosenza, Italy}

\date{\today}
\maketitle
\widetext

\begin{abstract}
\begin{center}

\parbox{14cm}{A Berry phase can be added  to  the wavefunction
of an isolated quantum dot by adiabatically modulating  a nonuniform
 orthogonal electric field,
 along a time-cycle. The dot is tuned close to
a three-level degeneracy, which provides a wide range of
possibilities of control. We propose  to detect  the accumulated phase
 by capacitively coupling the dot to a double-path interferometer.
The effective Hamiltonian  for the phase sensitive coupling is discussed 
in detail.}

\end{center}
\end{abstract}

\pacs{
\hspace{1.9cm}
PACS numbers: 03.65.Vf, 73.63.-b, 73.21.La
}
]

\narrowtext

\section{Introduction} 

Manipulating in a controlled fashion the phase of a quantum electronic system
 is presently one of the most relevant challenges in nanophysics, especially
in view of possible applications to quantum computing \cite{divincenzo}.
Probably, the most promising route to achieve such a task
is provided by coherent solid-state devices. For instance,
a superconducting Josephson qubit has already been realized  as a Cooper pair
box, that is,  a small superconducting island, weakly coupled to a charge
reservoir  via a Josephson junction  \cite{nakamura}. The quantum state of the
box can be tuned to a coherent superposition of the charge-zero
and the charge-one states.
The possibility of realizing superpositions of flux states has been 
considered, as well \cite{flux}. Entanglement in a semiconducting device
made out of two dots, one on top of each other (``quantum dot molecule'')
has recently been optically measured \cite{bayer}.

Usually, quantum  algorithms either assume that the system
dynamically evolves  through a sequence of unitary transformations, or
that a set $\vec{\lambda}$ of external control
parameters of the Hamiltonian $H$ smoothly changes in time   
(``adiabatic evolution'') \cite{farhi}. In
particular, if adiabatic evolution is realized across a closed path $\gamma$
in parameter space, close enough to an
accidental level degeneracy, the nontrivial topology of the space
makes the state of the system to take a ``geometrical'' phase $\Gamma$, 
referred to as ``Berry Phase'' \cite{berry}. The value of $\Gamma$
may be controlled by properly choosing $\gamma$.

 Following this idea,
geometric adiabatic evolution has recently been proposed as a way
to operate with superconducting devices without destroying phase coherence
\cite{shenoy,fazio}.
Another possibility is using semiconducting nanodevices, 
like single-electron  transistors  or
Quantum Dots (QD). The QD state can be finely tuned, by means of external
magnetic and electric fields acting on the dot, or on the coupling between
the dot and the contacts \cite{kastner}.
Moreover, accidental level degeneracies are quite common
in QD's, as seen both theoretically, and experimentally
\cite{benoit}.
Also, double dots have been proposed as possible qubits \cite{loss,pazy}.

\begin{figure}
\includegraphics*[width=0.87\linewidth]{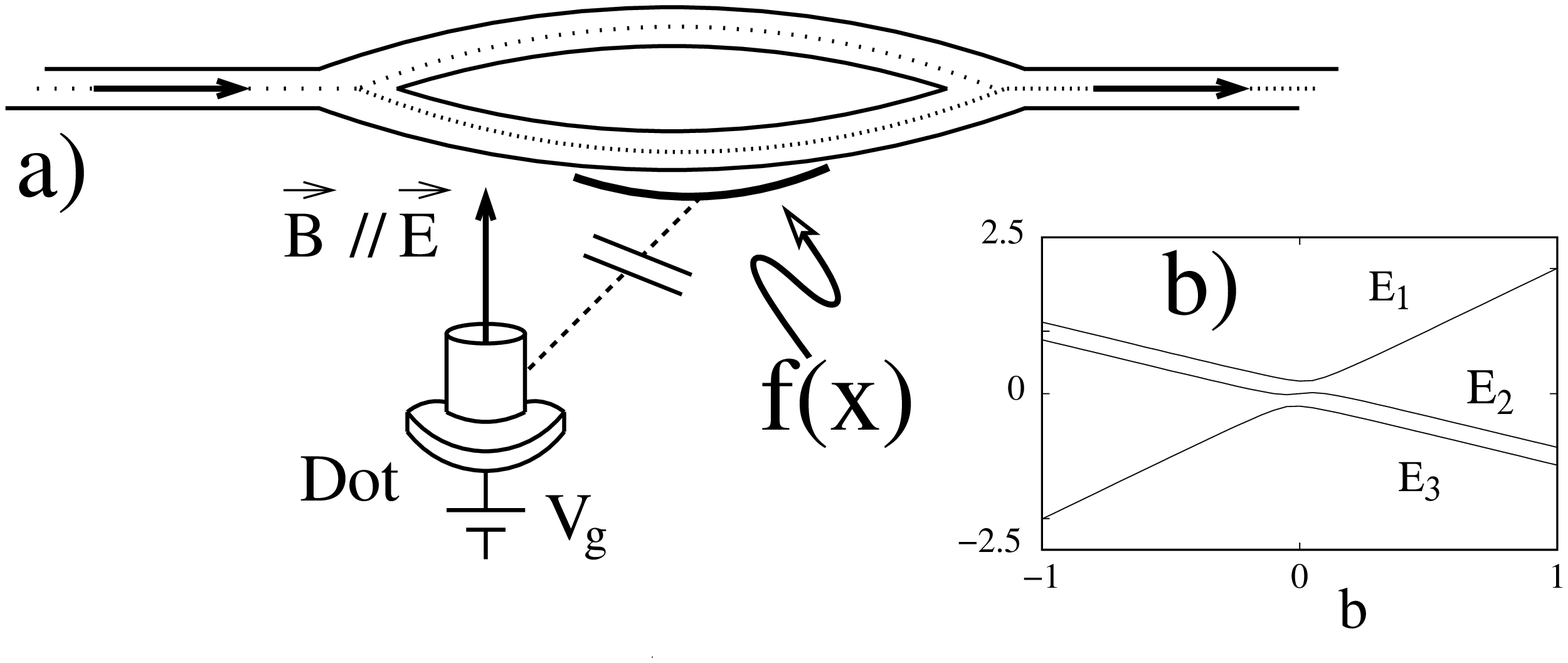}
\caption{{\bf a)}: A sketch of a possible experimental setup to detect the
Berry phase. $f(x)$ accounts for spatial modulation of the coupling
between the dot and the interferometer (see text).
{\bf b)} Energy levels of the dot, $E_\ell$ vs. $b=3/2 ( B-B^*)$.}
\label{fig1}
\end{figure}

In this paper, we address in its basic points the problem of adiabatic 
control in a single dot, that is, $a) $:  how to  generate   a Berry phase 
in an isolated dot (or in a QD at Coulomb Blockade);  $ b) $:  how to 
detect it. 

 $a)\: $ We focus onto a possible realization of the nontrivial 
topological structure studied in Ref.\cite{beppe2}, but  we  leave aside 
questions concerning the details of  the experimental device.
 After a time cycle, $\Gamma$ will show up as the phase factor of 
the dot ground state (GS).  We consider  an isolated  vertical QD  in  
vertical magnetic and electric fields,  $B$ and ${\cal E}$, respectively.
$B$ and ${\cal E}$ work as (adiabatic) control parameters of the
system. In particular, $B$ is tuned close to a two-level degeneracy,
taking place at $B=B^*$. The adiabatic cycle is realized by keeping $B$
fixed, and by slowly periodically varying ${\cal E}$, with time period $T$. 
Provided ${\cal E}$ is nonuniform over the dot's area, a spin-orbit coupling
term arises, involving the spin of electrons at the dot. Such an interaction
may give raise to a Berry phase at the QD. 
In particular, we will find a nonzero $\Gamma$ only
if $B > B^*$, that is, only on one side of values of $B$, with respect to the
avoided crossing point,  $B = B^*$.  As we will discuss in the paper,
this  is related to
the particular form of the adiabatic dot Hamiltonian, $H_D$, we operate
with, that is different from the usual $su(2)$-spin Hamiltonian introduced by 
Berry \cite{berry}.

Our model realization of the Berry phase in a quantum dot is taken as the
simplest setup that can be theoretically studied, with the required
features. Actually, its   experimental realization  is quite demanding, 
at least as long as one is concerned with a single dot only. 
For instance, one may think of a setup where the maximum of 
${\cal{E}}$  could  be off-center in the 
dot area and rotating in time. Alternatively, slowly time-dependent 
asymmetries in the shape of the dot might produce a Berry phase.

Very likely, better chances of realizing adiabatic evolution, based on
the same principles, may be achieved in molecular structures of strongly 
interacting dots \cite{bayer}.

$b)$ In order to measure the phase at the dot, we propose an experimental
arrangement, borrowed from an analogous experiment
\cite{ringarm}. We imagine to capacitively couple the dot to one arm of a 
double-path electron  interferometer (Fig.(\ref{fig1}a)). We show how the 
phase carried by the transported electrons may be influenced  by the dot.
The dot's phase gives raise to an interference term in the total conductance
across the ring. 

In Appendix B, we discuss the features of the coupling between the dot and 
the arm of the ring, which contributes to an effective Hamiltonian 
for the conduction electrons in the arm coupled to the dot 
(hereafter referred to as ``lead electrons'', as well). Such a coupling
should be weak enough, not to affect the modulus of the electronic 
transmission amplitude. 
We show that  the phase sensitive terms are, in general, nonlocal, both in
time and in space, and that they only arise if the capacitive coupling 
is extended in space on a length $L \sim  v_F T $, where $v_F $ is the Fermi 
velocity of the electrons  in the metallic lead.

In Appendix A, we derive the extra phase $\Phi(t,t')$ entering 
the propagator of the charge  density of the dot, $G_q(t,t' )$, which 
appears in  the lead  electron Hamiltonian.  Because of the 
adiabatic cycling, time-translational invariance of  $G_q(t,t' )$ is lost. 
 We compute the phase $\Phi$,  as well as $\Gamma$,  and analyze in
detail their expressions. A closed expression for  $\Phi (t,t')$ can only 
be given for $t\to t' \to 0 $.   
We find that, while $\Gamma $ is determined only  by  the geometrical 
properties  of $H_D$,  $\Phi $ (given in Eq.\ref{faso}), is nonlocal
in time, and contains  dynamical factors, that  
cannot be disentangled from the geometrical ones.

Finally, we argue that  the  essential features of the phase dependence are 
contained in the  time dependent model Hamiltonian  $ H_W ^{eff}$ for
lead electrons, given in Eq.(\ref{tham}). 
Using the simplified Hamiltonian $H_W ^{eff} $ allows us to straightforwardly 
derive the effects of the extra phase arising 
at the dot on the transmittance of the lead electrons.

In conclusion, we expect that, by steadily cycling the  dot Hamiltonian along 
a given path in parameter space,  a phase difference sets in between the 
electrons  of the two arms of the ring. The  phase picked up by the 
electrons in moving along the arm facing the dot is  averaged over a time
interval $L / v_F$, and should be measurable as an interference contribution to
the total conductance across the ring. 

The  very special properties  of the Berry phase that we produce, 
allow us to discriminate whether we are measuring a phase sensitive 
effect or just an unwanted electrostatic influence of the dot on the 
conductance in the ring. Indeed, the same operation on the dot can be 
performed at $B > B^* $ or $B< B^* $. However, the case  
$B< B^* $ does not provide the lead electrons with a Berry phase.
If we are really acting on the phase of the lead 
electrons  and not on the modulus of their transmission, the interference
should appear in the former case, but not in the latter. 

The paper is organized as follows: 

\begin{itemize}

\item in Section II we discuss the model Hamiltonian for the quantum dot, 
$H_D$. In particular, we specify the level structure and the control parameters
of the dot. Having chosen a subspace of states appropriate to the range of 
parameters we are considering, we derive the corresponding 
eigenvalues, as functions of the tuning parameters;

\item in Section III we calculate the Berry phase for the isolated dot. We show
that, as the applied magnetic field $B$ crosses  $B^*$, 
there is an abrupt jump  in the Berry phase from 0 to a value of order $\pi/2$.
Moreover, we also show that, for $B> B^*$ the Berry phase is  largely  
insensitive of $B$;

\item in Section IV
  we show how the Berry phase arising at the dot may be 
picked up by the  wavefunctions of the conduction electrons in one of the 
arms of   a ring interferometer.  Using results about the adiabatic evolution 
in the dot, derived in Appendix A, we obtain 
an effective Hamiltonian for the lead electrons, including  phase
dependent terms  (the details are reported in Appendix B); 

\item in Section V we discuss our results.

\end{itemize}

\section{The Dot  Hamiltonian $ H_D$ }

The dot's Hamiltonian, $H_D$,  
adiabatically depends on the external  parameters $B$ and $\cal{E}$, 
generically referred to as $\vec{\lambda}$, in the following.
In this Section we  introduce $H_D$ and  discuss how its
  eigenvalues and eigenstates depend on $\vec{\lambda}$.

To be specific, we consider an  isolated, vertical QD, disk-shaped in 
the $(x,y)$-plane. An external static magnetic field $B$ and an electric 
field ${\cal E}$ are applied along the  $z$-axis. ${\cal E}$ takes a small 
angular modulation: ${\cal E} = e + g_1 \cos ( \theta ) + g_2 \sin  (\theta)$.
($\theta$ and $\rho $ are the  polar coordinates  in the 
plane). The corresponding dot Hamiltonian is :

\beq
 H_{D} = H_{0D} + \sum_i 
[ \alpha ^2  ( \vec{\cal E}\times \vec{p}_i )\cdot
\vec{\sigma}_i  + \mu  l^z_i B ] + U(B, \omega _d)
\label{ham}
\eneq
\noindent
where $H_{0D}$ includes  the confining parabolic potential
of  frequency $\omega_d $, the $\vec{p} _i$'s and the $\vec{l}_i$'s are the
linear and the angular momenta of the electrons at the dot, respectively.
The second and third terms at the r.h.s. of Eq.(\ref{ham}) are  the
spin-orbit (Rashba) term  and the Zeeman term, respectively.
$\alpha^2$ is the spin-orbit
coupling constant and  $\mu$ is
the electronic magnetic moment.  $U(B, \omega _d ) $ is the 
operator  for the Coulomb interaction.
We neglect  the Zeeman spin splitting, which is expected to be small.

The states of a vertical dot are usually denoted as
$ | N , M , S , S^z \rangle$, where $M$ is the orbital angular momentum, $S$
is the total spin and $S^z$ is its $z$-component \cite{benoit}.
The dot state may be controlled by properly
tuning the external control parameters  $B$ and $\cal{E}$.

For $N=3$, the unpaired electron is the only one to be ``active'', in a wide 
range of parameters. Therefore, we may label the dot's states with the quantum
number of such a ``valence'' electron only, i.e. 
$|n,m, s=\met , s^z \rangle $ ($s, s^z$ are the electron spin and its 
$z$-component). Also, $n,m$ are the orbital quantum numbers, corresponding to 
an orbital wavefunction for an electron in a harmonic confining potential and 
an external $B$ field, given by:
\beq
\Psi_{n,m} ( \rho, \theta ) =  
\frac{e^{im\theta}}{l \sqrt{\pi}}  \: R _{n|m| } (t)
\label{adj1}
\eneq
\noindent
 with   $ t = \rho ^2 / l^2 $,  where  $l = \sqrt{ \hbar / m \omega _0} $ 
 ($\omega _0 = \sqrt {\omega _d^2 + \omega _c^2(B) /4}$
and $\omega _c(B)$  is the cyclotron frequency). 
 The radial wavefunction in Eq.(\ref{adj1}) is expressed in terms of 
the Laguerre 
polynomials $ L^\mu_\nu $ as:
\beq
R_{n|m| } (t) =  {\cal{C}}_{n|m|}  \: e^{-\frac{t}{2 } } \:  t^{|m|/2} \:
L^{|m|} _{\frac{n-{|m|} }{2}}  \: (t )  
\eneq
\noindent
where  $ {\cal{C}}_{n|m|}  = \left [ 
\frac{\left ({\frac{n-{|m|} }{2}}\right ) !}
{\left ({\frac{n+{|m|} }{2}}\right ) !} \right ]^\met $.

Numerical diagonalization of Eq.(\ref{ham}) shows that, at 
 $B=B^*$ and ${\cal E}=0$, the states 
$ | 1 ,  \vec{\lambda} \rangle \equiv|1  1 ,  \frac{1}{2} , - \frac{1}{2}, 
\vl \rangle$ and $| 3 , \vec{\lambda} \rangle \equiv | 2 2 , \frac{1}{2} , 
-\frac{1}{2}, \vl \rangle$ 
become degenerate with $ | 2, \vl \rangle \equiv  | 1 1 , \frac{1}{2} , 
\frac{1}{2}, \vl \rangle$, because of the $e-e $ interaction 
\cite{unpublished} (we add $ \vec{\lambda} $ in the notation to stress 
that they depend  on the value of the external parameters).

Other  levels are much higher in energy, so that here we employ a 
3$\times$3-Hamiltonian matrix, to diagonalize the spin-orbit term.

At ${\cal E} \neq 0$,  the spin-orbit  term couples  states with opposite 
spin components:
$ s^z = \pm \met $. The  isotropic component of $\cal{E}$ has matrix elements 
between states  with  $J= M + S^z = 3/2 $ (i.e.  the orbital state 
$|3 , \vec{\lambda} \rangle $,  and  the orbital state 
$|2, \vec{\lambda}  \rangle $). The matrix elements of the   anisotropic terms 
($ \propto \:\rho \:\sin \theta $ and $  \propto  \:\rho \:\cos \theta $)
can be easily calculated. For instance, one obtains:
\begin{eqnarray}
A_{n'm'+, \: n m -} = \langle n'm'|  \:\rho  \:\sin \theta  
 e ^{-i \theta } (\partial _\rho  +\frac{m}{\rho} ) |n m \rangle\nonumber\\
 = \frac{1}{ i } ( \delta _{m' +2 , m} - \delta _{m', m} )\nonumber\\
 \times \met \int _0 ^\infty   \sqrt{t} \: dt \: R_{n'|m'| } (t)
\left (2\sqrt{t} \partial _t  +\frac{m}{\sqrt{t}}
\right  ) R_{n|m| } (t)
\label{matr}
\end{eqnarray}
\noindent

From Eq.(\ref{matr}) we see  that $ | 2 , \vec{\lambda} \rangle$ is coupled to 
$ | 1 , \vec{\lambda} \rangle$, with matrix element $ g = g_1 - i g_2$.  

Therefore, for $B$ close to $B^*$, once restricted to the
subspace spanned by the states $ | \ell , \vl \rangle $ ($\ell = 1,2,3$), 
$H_{D}$ can be  represented by the matrix $\hat{h}$:

\beq
\hat{h} [ b , g_1, g_2 , e ] =
\left[ \begin{array}{ccc} - b & 0 & g^* \\ 0 & 2 b & e \\ g & e & - b
\end{array} \right] \: ,
\label{eqadone}
\eneq
\noindent
where  $ b= 3(B-B^*)/2 $.

$\hat{h} [ b , g_1, g_2 , e ]$ is a traceless 3$\times$3 Hermitian matrix,
belonging to the $su(3)$-algebra. Its eigenvalues take
a simple form in terms of the ``polar'' coordinates $R , \Psi$, defined by:

\[
R = \sqrt{b^2 + \frac{e^2 + |g|^2}{3}} \;\;\; ; \;
\sin ( 3 \Psi ) = - \frac{b ( b^2 + \frac{e^2}{2} - |g|^2 )}{ R^3}
\]
\noindent

In decreasing order, the energies are given by \cite{beppe2}:

\beq
E_\ell = 2 R \sin \left[ \Psi + \frac{2}{3} ( \ell - 1 ) \pi \right]
\;\;\; \ell =1,2,3
\label{eqadtwo}
\eneq
\noindent

 In Fig.(\ref{fig1}b), we plot $E_1,E_2$ and $E_3$ versus $b$, 
at $e$, $g$ small, but $\neq 0$.
The corresponding eigenvectors  will be denoted 
 in the following   by  $ | e_\ell , \vec{\lambda} \rangle$, 
$\ell = 1 , 2 , 3 $.

 Fig.(\ref{fig1}b) shows the avoided crossing  vs. $b$. In particular, if  
$b < 0$  the level $E_1$ is almost degenerate with $E_2$, and the degeneracy 
at $e=g=0$ takes place at $\Psi = \pi / 6$. On the other hand,  if $b>0$,  
it is $E_3$ to be almost degenerate with $E_2$, and the degeneracy takes place 
at $\Psi=\pi/2$. At $b=0$  an ``exceptional'' three-level degeneracy arises, 
when $e=g=0$ ($R=0$), although, as discussed in detail in the next Section,
 this is not  an accidental three-level degeneracy. 

In the next Section, we show that  since, for $b<0$, there is an energy gap 
of order $b$ between $E_3$ and
the next available energy level ($E_2$), we do not expect any Berry phase to
arise within such a region. On the other hand, we do expect a Berry phase to
appear for $b>0$, when $E_3$ keeps almost degenerate with $E_2$. Therefore, 
whether a Berry phase may arise, or not, is just matter of whether $b$ is
$> 0$, or $< 0$. Clearly,  no fine tuning of the external field is
required, provided it is possible to move $B$ across $B^*$.

\section{Calculation of the Berry phase}

In this Section we calculate the Berry phase accumulated at the quantum dot,
by adiabatically operating along a cycle periodic in time, with period $T$.  
We will consider a closed path $\gamma $, lying in the subspace of 
coordinates $(g_1 , g_2 , e )$, at fixed $b$.
At $b \neq 0$, a two-level accidental degeneracy  occurs when $e =g=0$.
It is possible to obtain a nonzero Berry phase if $\gamma $ encircles this 
point. 

When dealing with such a kind of problems, it is customary to define
a one-parameter family of Hamiltonians, $H_A ( s)$, such that
$ H_A ( s ) = H_{D} [ \vec{\lambda} (  s ) ] $, $\forall s$
($s = t /T$,  $s \in [ 0 , 1 ]$).

The solution of the time-independent parametric Schr\"odinger equation
associated to $H_A ( s)$ provides the eigenvalues $E_l [\vec{\lambda} (s)]$ and
the corresponding normalized eigenvectors $ | e_l, \vec{\lambda} ( s ) 
\rangle$  (also denoted by  $E_l (s)$ and $ | e_l ( s ) \rangle$, in the
following).

Let $| F(t) \rangle$ be the ground state of the system. At $t=0$ we have
$| F ( 0 ) \rangle = | e_3 ( 0 ) \rangle$, that is, the ground state
coincides with the ground state of the adiabatic Hamiltonian.  Moreover, 
adiabaticity  implies $ | F ( T ) \rangle = e^{ i \Gamma} 
| e_3 ( 0 ) \rangle$, where the geometrical phase $\Gamma$ is given by:

\beq
\Gamma = i \oint_\gamma  d \vec{ \lambda} \cdot \;\biggl\langle
e_3, \vec{\lambda}   \biggl|  \vec{\nabla}_{\vec{\lambda}}
\biggr|  e_3 , \vec{\lambda}    \biggr\rangle
\label{piuone}
\eneq
\noindent

By means of Stoke's theorem, $\Gamma$ may also be written as a two-dimensional
integral on a surface  ${\cal S}$, bounded by $\gamma$ \cite{berry}:

\begin{eqnarray}
\Gamma = \int_{\cal S}  \; (  d \lambda ^a \wedge d \lambda ^b \; ) \nonumber\\
\times 
\sum_{\ell \neq 3 } \; {\Im} m \biggl\{ \frac{ \langle
e_3 , \vec{\lambda}  | 
\frac{ \partial H}{ \partial \lambda ^a} | e_\ell,\vl \rangle \langle 
e_\ell,\vl | \frac{ \partial H}{\partial \lambda ^b} |  e_3 ,\vl \rangle }{
( E_\ell [\vec\lambda] - E_3 [\vec\lambda] )^2} \biggr\}
\label{eqtenbis}
\end{eqnarray}
\noindent
where  $d \lambda^a \wedge d \lambda^b$ is the
projection of the surface element of ${\cal S}$ on the  $( a,b )$ plane in
parameter space.

\begin{figure}
\includegraphics*[width=0.87\linewidth]{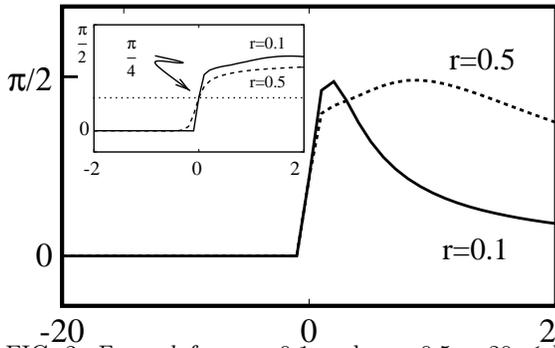}
\caption{$\Gamma$ vs. $b$ for $r=0.1$ and $r=0.5$, $-20 \leq b \leq 20$.
{\bf Inset}: Zoom of the plot for  $-2 \leq b \leq 2$.}
\label{fig2}
\end{figure}

We have explicitly calculated $\Gamma$,
in the case where $\gamma $ is the closed path on the sphere of radius $r$ 
defined by $g= r \sin ( \vartheta ) e^{ i \varphi}$, $e = r \cos 
( \vartheta )$, $ \vartheta = \pi / 4$ (the detailed shape of $\gamma$ is, of
course, irrelevant to our calculation, and we have chosen a  circular path 
at  $ \vartheta = \pi / 4$ just for the sake of simplicity).
 We have numerically computed $\Gamma$ at 
fixed $b$, $\Gamma ( b )$, by using Eq.(\ref{eqtenbis}), where, in our case,
${\cal S}$ is identified with the spherical segment ${\cal S} : 0 \leq 
\vartheta \leq \pi / 4 ; 0 \leq \varphi \leq 2 \pi$. The derivatives of
$\hat{h}$ with respect to the various parameters that matter for the
calculation of $\Gamma$ are provided by the following $su(3)$-generators:

\[
\frac{ \partial \hat{h}}{\partial g_1} = \left[ \begin{array}{ccc} 0 & 0 & 1\\
0 & 0 & 0 \\ 1 & 0 & 0 \end{array} \right] \;\;
\frac{ \partial \hat{h}}{\partial g_2} = \left[ \begin{array}{ccc} 0 &0&- i\\
0 & 0 & 0 \\ i & 0 & 0 \end{array} \right] 
\]

\[
\frac{ \partial \hat{h}}{\partial e } = \left[ \begin{array}{ccc} 0 & 0 & 0\\
0 & 0 & 1 \\ 0 & 1 & 0 \end{array} \right] 
\]
\noindent
In  Fig.(\ref{fig2}) we plot  $\Gamma ( b )$ vs. $b$. We see that, while
$\Gamma$ is close to zero, as long as $b<0$, it abruptly jumps to a finite
value (of order $\pi/2$) across  $b=0$. As $b>0$, it keeps rather flat, till
it starts to decrease for decreasing $r/b $. The extremely weak dependence of
$\Gamma$ on $r/b$ for $b>0$ is quite important for the practical purpose of
detecting the Berry phase. Indeed, it is the very reason why a fine tuning
of $b$ is not required, in order to measure $\Gamma$, provided $b>0$.

In order to explain the plot in Fig.(\ref{fig2}) and, in particular, the 
behavior of $\Gamma ( b )$ for $b >0$, we have gone through an approximate 
analytical calculation of the Berry phase, which we are now going to 
discuss. 

The infinitesimal contribution to the Berry phase for the system lying within 
the ground state of  energy $E_3$, $d \Gamma$, is:

\begin{eqnarray}
d \Gamma \approx {\Im}m \sum_{\ell =1,2} \frac{1}{ (E_3 - E_\ell )^2 } \times
\nonumber\\
\biggl\{ d g_1 \wedge d g_2 
\langle e_3 , \vec{\lambda} | \frac{ \partial \hat{h}}{\partial g_1} | 
e_\ell , \vec{\lambda} \rangle \langle e_\ell , \vec{\lambda}  
| \frac{ \partial \hat{h}}{ \partial g_2} | e_3 , \vec{\lambda} \rangle 
\nonumber\\
+ d g_2 \wedge d e 
\langle 3 , \vec{\lambda} | \frac{ \partial \hat{h}}{\partial g_2 } | 
e_\ell , \vec{\lambda} \rangle \langle e_\ell , \vec{\lambda}  
| \frac{ \partial \hat{h} }{  \partial e }| e_3 , \vec{\lambda} \rangle 
\nonumber\\
+ d e \wedge d g_1 
\langle e_3 , \vec{\lambda} | \frac{ \partial \hat{h}}{\partial e } | 
e_\ell , \vec{\lambda} \rangle \langle e_\ell , \vec{\lambda}  | \frac{ 
\partial \hat{h} }{ \partial g_1} | e_3 , \vec{\lambda} \rangle \biggr\} \;\; .
\label{addb3}
\end{eqnarray}
\noindent

Let us separately analyze the two complementary regimes $b<0$ and
$b>0$.

\begin{itemize}

\item {The case $b<0$:}

In this regime, up to ${\cal O} ( r^2 / b^2 )$, we obtain the following
approximate expressions for the energy eigenvalues:

\begin{eqnarray}
E_1 \approx - b + | g | - \frac{ e^2}{ 6 b}\nonumber\\
E_2 \approx - b - | g | - \frac{ e^2}{ 6 b}\nonumber\\
\nonumber\\
E_3 \approx 2 b + \frac{  e^2}{ 3 b } 
\label{addb1}
\end{eqnarray}
\noindent

Since $(E_3 - E_\ell )^2 \sim b^2$, it is straightforward to derive that, up to
${\cal O} (r^2 / b^2 )$, one obtains $d \Gamma_< = 0$ and, therefore, $\Gamma
(b) = 0$, as long as $b<0$. Such a strong suppression of the Berry phase is
clearly seen from our exact numerical diagonalization results, and may be
understood as a consequence of the absence, for $b<0$, of energy levels near 
by $E_3$.

\item{The case $b>0$:}

In this case, the energy eigenvalues are given by:
\begin{eqnarray}
E_1 \approx 2 b + \frac{e^2}{3 b}\nonumber\\
E_2 \approx - b + |g| - \frac{e^2}{ 6 b }\nonumber\\
E_3 \approx - b - | g | - \frac{e^2}{ 6 b }
\label{addb2}
\end{eqnarray}
\noindent

Since $E_2 - E_3 $ is ${\cal O} ( | g |)$, we may calculate $\Gamma$ by taking
into account only these two levels. The direct calculation of the various
matrix elements provides us with the result:

\[
{\Im}m \biggl\{ 
\langle e_3 , \vec{\lambda} | \frac{ \partial \hat{h}}{ \partial g_1}
| e_2 , \vec{\lambda } \rangle \langle e_2 , \vec{\lambda } | \frac{ \partial}{
\partial g_2} | e_3 , \vec{\lambda} \rangle \biggr\}
= - \frac{e^2}{ 6 b | g |}
\]

\[
{\Im}m \biggl\{ 
\langle e_3 , \vec{\lambda} | \frac{ \partial \hat{h}}{ \partial g_2}
| e_2 , \vec{\lambda } \rangle \langle  e_2, \vec{\lambda } | \frac{ \partial}{
\partial e } | e_3 , \vec{\lambda} \rangle \biggr\}
=  - \frac{e g_1 }{ 3 b | g |}
\]

\beq
{\Im}m \biggl\{ 
\langle e_3 , \vec{\lambda} | \frac{ \partial \hat{h}}{ \partial e }
| e_2 , \vec{\lambda } \rangle \langle e_2 , \vec{\lambda } | \frac{ \partial}{
\partial g_1 } | e_3 , \vec{\lambda} \rangle \biggr\}
 =   \frac{e g_2 }{ 3 b | g |}
\label{addb4}
\eneq
\noindent
Since $ (E_3 - E_2 )^2 \approx 4 |g|^2$, we obtain the following expression
for $d \Gamma_>$:

\beq
d \Gamma_> = - \frac{r}{3b} \biggl[ 1 - \frac{1}{ 2 \sin^2 ( \vartheta )
+ a r^2 / b^2 } \biggr]
\cos ( \vartheta ) d \vartheta \wedge d \varphi
\label{addb5}
\eneq
\noindent
where $a$ is a numerical constant that is ${\cal O} (1)$.
\end{itemize}

Notice that $d \Gamma_>$ is given by two contributions. The first one may be
clearly identified with the usual $su(2)$ elementary Berry phase \cite{berry}.
Indeed  by introducing the polar coordinate $\vartheta^{'} = \vartheta - 
\pi /2$, it takes the form:

\[
d \Gamma_>^{(1)} = \frac{ r}{ 3 b } \sin ( \vartheta^{'} ) d \vartheta^{'} 
\wedge d \varphi
\]
\noindent
where the ``monopole strength'' is given by $  4 \pi r / 3 b$. Once
integrated, $d \Gamma_>^{(1)}$ provides the contribution to the 
finite Berry phase given by:

\beq
\Gamma^{(1)} = - \frac{ \sqrt{2} \pi r}{3b}
\label{addb6}
\eneq
\noindent
that is suppressed as $r/b$, for large $b/r$.

The second contribution, on the other hand is quite peculiar of the 
specific problem we are dealing with, and does not take any resemblance with
the usual $su(2)$ Berry phase. We have traded higher-order term appearing
at the denominator for a cutoff of order of $r^2 / b^2$ that, once integrated,
provides us with the additional contribution to $\Gamma$ given by:

\beq
\Gamma^{(2)} = \frac{ \pi }{ 3 \sqrt{a}} {\rm atan} \left( \frac{ \pi b }{ 4 r}
\right)
\label{addb7}
\eneq
\noindent
Already for $b / r \sim 5$ we may approximate $\Gamma^{(2)}$ as $ 
\approx \pi^2 / 6 \sqrt{a}$. The deviation of $\Gamma (b)$ from a constant
value, for $b > 0$, which may be clearly seen from Fig.(\ref{fig2}), is
due to contributions that we have neglected in writing Eq.(\ref{addb5}). 
Nevertheless, our result
clearly shows how the peculiar form of the Berry phase in our problem makes it
possible to have a finite and detectable $\Gamma$ even for $r / b \sim 0.005$.
Again, let us remark that this result is quite important, since it shows that
no fine-tuning of $B$ is needed, in order to give raise to a Berry phase,
provided one is performing the measurement in the region $B > B_*$.

It is also possible to exactly evaluate
the Berry phase at  $b=0$. This is  an ``exceptional'' point, where
$e=g=0$ implies a three-fold degeneracy among all the energy levels,
corresponding to $\Psi = \pi / 3$. Such a degeneracy takes place among
three states, two of which differ by the spin polarization of the unpaired
electron. Hence, it is not appropriate, here, to speak of an ``accidental
three-level degeneracy'', since two of the three levels involved are
connected by means of a continuous $SU (2)$-symmetry. Thus, the three-level
degeneracy we find in a four-parameter Hamiltonian is not inconsistent with
the result of Ref.\cite{berry}, where it is stated that one needs
at least eight parameters, in order to get an accidental three-level 
degeneracy with a complex Hamiltonian\cite{referee}. 

Close to the three-level degeneracy, we obtain:

\[
\Gamma ( b = 0 ) =  \int_{\cal S} \frac{1}{ 4 r^4}
\{ e^2 d g_1 \wedge d g_2 +  e g_2 d e \wedge d g_1  + e g_1 d g_2 \wedge
d e \}
\]

\beq
=  \frac{1}{2}
\int_0^{ 2 \pi} d \varphi \int_0^{\frac{\pi}{4}}  \sin ( 2 \vartheta )
d \vartheta = \frac{\pi}{4} \;\;\; ,
\label{eqadfive}
\eneq
\noindent
independently of the radius of the sphere.

In the inset of Fig.(\ref{fig2}) we show a zoom of the region around $b = 0$.
Curves  for different radii intersect at $b=0$ with a value equal to $\pi / 4$,
as derived in Eq.(\ref{eqadfive}).

\section{Detection of the Berry phase}

In Section III we have calculated the total Berry phase for the isolated
dot $\Gamma$.  We have shown that, for the special  time dependent Hamiltonian
 we consider, the adiabatic cycling  over a period $T$ gives $\Gamma \neq 
0 $ or $ =0 $  depending on whether $b>0  $  or  $<0  $.
 
While one usually looks at the effects
of the adiabatic cycling on the wavefunction of the isolated system (the 
Berry phase of the dot, in our case), in this work  we want to study how 
the adiabatic evolution of the dot may affect the propagator for 
conduction electrons traveling in some conductor out of the dot,
  but weakly interacting with it. 
Therefore, our  second goal of our work  is to compute the relation between
 the electron propagator  of the dot in the presence of the  Berry phase 
cycling and the adiabatic one. This is done in  Appendix A
  for the charge propagator:
\beq
G_q ( t, t^{'} ) =
{\rm Tr}  \left [ e^{-\beta H_D }\:{\cal {T}} 
\: U( T,t) \:   \hat{q} (t) \hat{q} (t') \: U(t',T ) \right ] \:  .
\label{gq}
\eneq
Here ${\cal {T}}$  is the time ordering operator and 
 $U ( t, t^{'}  )$ is the time-evolution operator corresponding to the
 dot Hamiltonian $ H_D ( t )$. It is given by: $U (t,t^{'} ) =
{\cal T} \exp [ - i \int_{t^{'}}^t d \tau H _D ( \tau ) ]$.
Such a calculation requires the implementation of  
non equilibrium Green's function approach. In Appendix A, we present
such an approach. In particular, we make use of the adiabatic evolution in
the QD dynamics and show that the difference between the full
and the adiabatic Hamiltonian is ${\cal{O}} (1/T )$.

The  phase factors containing  $\Phi (t,t')$,  which relate the adiabatic 
Green's  function to the full single particle propagator are computed 
 explicitely in Appendix A, only in the limit  $t\to t' $.  By comparing 
the derivation  for $\Phi $  with the expression of   $\Gamma $ 
given by  Eq.(\ref{eqtenbis}),
we see that it is not possible to reduce the latter to the former, in any
straightforward limit.
 Indeed, $\Gamma$ is determined by geometrical 
factors only, while $\Phi $  contains also dynamical phase factors, 
which it is impossible to disentangle from the geometrical ones, 
due to nonlocal contributions  ( see e.g. Eq.(\ref{faso})).
However, the derivation shows that   $\Phi $ vanishes, if $\Gamma $ does so.
It is easily seen that  that   the phase  $\Phi _o $
of Eq.(\ref{faso})  and $\Gamma $ coincide only when 
the nonlocal terms are dropped.

 Our  third  task is to find out if it is possible to detect the phase 
$\Phi $  by means of  a  conductance measurement.

 In this Section,  we propose an experimental way
 to detect a Berry phase in the dot wavefunction.
We consider a double-path electron interferometer, weakly interacting with  the
dot by means of a capacitive coupling (see Fig.(\ref{fig1})). In such a kind 
of experimental arrangement, electrons passing
through the arm of the interferometer coupled to the dot may pick up a
finite phase  $\Phi (t,t')$. Meanwhile, 
a fine tuning of the coupling should keep the  modulus of 
the transmission of each arm roughly  unitary. We show here that 
if the  coupling 
between the dot and the interferometer is dealt with perturbatively,
 the phase $\Phi (t,t') $  depends on the adiabatic dynamics of the 
isolated dot only, that is on the Green's function of Eq.(\ref{gq}).

 In order to do so,  we derive  an effective Hamiltonian for
lead electrons, when the interferometer  arm is capacitively 
coupled to  the dot, by integrating out the dot degrees of freedom.
Since, with a point-like contact, it is impossible to effectively tranfer
the phase from the dot to the interferometer,  we assume that the 
capacitive contact is extended over  a length $L$, that is,
 the support of an  envelope function $f(x) $
 (Fig.($1a$)). This gives raise to a nonlocal interaction among
dot electrons. We also assume conduction electrons to propagate 
ballistically in the wire. 

Using the results of Appendix B, we show that the nonlocal term responsible
for transferring the phase from the dot to the lead can be modelized by 
a local effective Hamiltonian $ H^{eff}_W $ given by  Eq.(\ref{tham}). 
The nonequilibrium, although adiabatic, dynamics of the system gives raise 
to a phase-dependent interference between wave components from different 
points in space, as it is evident in Appendix B,
where both the time ordered and the anti-time ordered Green's functions show 
up in the derivation. 

To deal with nonlocal interactions, we need to make use of the functional
integral approach.

Let $S_D [ \hat {c}(\underline{r}) ,\hat {c}^\dagger (\underline{r})  ]$
be the dot's action, derived from $H_D$  in Eq.(\ref{ham}),  in terms of the 
Fermionic fields of the dot electrons
$\hat {c}(\underline{r}) ,\hat {c}^\dagger (\underline{r}) $. Moreover, since
we model the arm interacting with the dot as a one-dimensional conductor, we
may expand the corresponding Fermionic field $\Psi (x)$ around the 
Fermi points $\pm q_F$, so that:

\beq
\Psi (x) = e^{ - i q_F x } \psi_L ( x ) + e^{ i q_F x } \psi_R ( x )
\label{lr}
\eneq
\noindent
where $\psi_{L,R} (x)$  are the left-handed  $(L)$ and the right-handed
 $(R)$ component, respectively, and $x$ is the coordinate along the wire. 
Since all the interactions, we are going to make use of, are 
diagonal in the spin index $\sigma$, we  just suppress it.
 The free  action of the wire is  given by:

\[
S_{W} = i \int d t \int d x \biggl[ \psi_L^\dagger ( x , t ) 
 \biggl( \frac{ \partial}{ \partial t} - v_F \frac{ \partial}{ \partial 
x} \biggr) \psi_R ( x , t )
\]

\beq
+ \psi_R^\dagger ( x , t )  \biggl( \frac{ \partial}{ \partial t} + 
v_F \frac{ \partial}{ \partial x} \biggr) \psi_R ( x , t ) \biggl]
\label{eis}
\eneq
\noindent
where $v_F = \hbar q_F / m \: ( \hbar =m=1 )$ is the Fermi velocity.

The capacitive coupling between the dot and the wire, depends on the charge 
density operator at the edge of the dot (parametrized by $\underline{r}_b $), 
$\hat{q}(\underline{r}_b)   =  \hat {c}^\dagger (\underline{r}_b)  
\hat {c}(\underline{r}_b) $.  
The coupling is assumed to smoothly extend
over a distance $ L \sim  v_F T $ and an envelope function  $ f( x)$  entails  the 
geometry of the gate  facing the  lead:
\beq
S_{\rm int} =  \int dt \:q (t) \: \int dx \: f( x )\: \Psi ^\dagger ( x , t )
\Psi ( x , t ) +   c.c. 
\eneq
(The space dependence of the  charge density $q$ is ignored because it plays 
no relevant role).

The effective propagator for electrons in the arm is given by:

\begin{eqnarray}
G(x ,t ; x^{'} ,t^{'}  )  = \int {\cal D}[ \Psi  \Psi ^\dagger  ] 
  \Psi ^\dagger (x ,t ) \Psi (x^{'} ,t^{'}  ) \:
 \nonumber\\ e^{ - i  S_{W} }
   \langle  e^{i\int dx\: f(x)\: \int \; dt \:q (t) \: 
  \Psi ^\dagger (x,t)  \Psi (x,t)}
\rangle _{ S_D}\;\;\; ,
\label{Z}
 \end{eqnarray}

with  appropriate  boundary conditions.

To lowest order in a cumulant expansion, the average 
$\langle \ldots \rangle _{ S_D}$  over the dot variables 
yields:
\begin{eqnarray}
   \langle  e^{i\int dx\: f(x) \:  \int  dt  q (t) \: 
  \Psi ^\dagger (x,t)  \Psi (x,t)}
\rangle _{ S_D} \approx \nonumber\\
  e^{-\met 
\int dt dt'\; G_q (t,t^{'})\int dx dx'\: f(x) f(x') \: 
  \Psi ^\dagger (x,t) \Psi (x,t)
  \Psi ^\dagger (x',t')  \Psi (x',t')}.\nonumber
\end{eqnarray}
where $G_q ( t, t^{'} ) $ is given by Eq.(\ref{gq})  and discussed in
 Appendix A.  

Eq.(\ref{Z})   shows that the capacitive coupling of the lead to the dot 
turns into an effective  interaction term  among  the lead 
electrons. The corresponding action comes out to be nonlocal, 
both in time and in space. It is so because the 
problem is intrinsically nonstationary \cite{keldysh}, although time 
dependence is assumed to be very slow.

We derive the  effective interaction  in detail in Appendix B.
 We first perform a mean field decoupling of 
the quartic term in the Fermionic fields in Eq.(\ref{Z}).
As a next step, we approximate the kernel to be local in time, but still 
nonlocal in space. In this way, it takes the form of 
an external potential, acting on the 
conduction electrons and giving raise to forward and backward scattering. 

As a consequence of the dot dynamics, one gets a 
nonlocal  potential, whose main effects may be summarized as 
follows:

$a) $ A $L-L$ and a $R-R$ electrostatic forward scattering potential, roughly
independent of the geometrical phase;

$b) $ A phase-dependent forward scattering term; 

$c) $ An electrostatic $L-R$ and $R-L$ backscattering potential, roughly 
independent of the phase;

$d) $ A phase-dependent $L-R$ and $R-L$ backscattering term.

As a further step, we ignore the  electrostatic perturbation  induced by 
the dot (terms $a) $ and $c) $ ). 
The reason for doing so is twofold. First of all, whatever such
a coupling is, its effects should not depend on whether a geometrical phase
is arising at the dot ($b>0$), or not ($b<0$), provided the path in parameter 
space is the same. Moreover, we assume that the 
corresponding couplings are weak enough, 
to make the various contributions negligible anyways (of course, this 
requires quite accurate experimental checks, but we do not want to deal 
with this issue  here).

On the other hand, terms $b) $ and $d) $ are 
 responsible for transferring the phase from the
dot to the lead. Such terms appear in a nonlocal form, though. These are 
the main results of the derivation reported in Appendix B. 

 The full Hamiltonian describing the capacitive coupling
  suggests that an effective local ``toy'' 
Hamiltonian $H_W^{eff}$ may be used, if  the effects of the Berry phase 
are the  main concern of our model construction. $H_W^{eff}$ 
is  given by:

\begin{eqnarray}
 H_W^{eff} &  = & H_0 + V_\Phi (t) + V_\chi (t)  \hspace*{1cm}
\label{tham}\\
H_0 & = & v_F  \int \frac{dk}{2\pi}\:k \left ( \psi ^{\dagger}_L (k)\psi _L (k) +
 \psi ^{\dagger}_R (k)\psi _R (k) \right )\nonumber \\
V_\Phi (t) & = &
 \dot \Phi (t)   \int _{-L/2}^{L/2}   dx \:
 \left ( \psi ^{\dagger}_L (x)\psi _L (x ) -  
 \psi ^{\dagger}_R (x)\psi _R (x) \right )
\nonumber\\
V_{\chi} (t)  & = &  v_\chi (t)  \int _{-L/2}^{L/2}  dx  \:
 \left ( \psi ^{\dagger}_L (x)\psi _R (x)  e^{2 i\Phi(t)} + \: h.c. \: \right )
\label{pot}
\end{eqnarray}

Notice that $H_W^{eff}$ in Eq.(\ref{tham}) explicitly depends on time.
 The time dependence is produced by  
adiabatically tuning the parameters of the dot. An expression for the 
coupling  $ v_\chi (t) $  can be derived  
from Eq.s (\ref{offdiag5},\ref{offdiag6})   within the approximations used 
in  Appendix B  and involves the retarded adiabatic Green's function, 
$  G_{q , A}^{\rm Ret} ( t^+ , t ) $, defined in Appendix A. In the 
``local approximation'' $\dot{\Phi}$ depends only on $t$, and is independent
of the space variables. 

The Hamiltonian $H_W^{eff}$  is particularly amenable because 
the phase $\Phi$   can be transferred onto the conduction electron wave 
functions  in the region $({-L/2}, {L/2} )$, by performing an unitary
time dependent transformation $e^{iS(t) }$ in such space interval, with  
\begin{eqnarray} 
S (t) =   \Phi (t)   \int _{-L/2}^{L/2}  dx  \:
 \left ( \psi ^{\dagger}_L (x)\psi _L (x ) -  
 \psi ^{\dagger}_R (x) \psi _R (x) \right ) .
\end{eqnarray}
Indeed, it is easy to show that
$e^{iS} \psi _L\: e^{-iS} = \psi _L  e^{-i  \Phi} \: $  
 and $ e^{iS} \psi _R\: e^{-iS} = \psi _R  e^{i \Phi} $,
so that
\begin{eqnarray}
     e^{iS}\{ \partial _t  -  \: H_{W}^{eff} \; \}   e^{-iS}  = \nonumber\\
  \partial _t - \left \{ H_0 +   v _\chi (t)  \int  _{-L/2}^{L/2}
  dx \:
 \left ( \psi ^{\dagger}_L (x)\psi _R (x)  + h.c.  \right ) \right \}
\nonumber\\
\equiv \partial _t -  H_{WA} \: . 
\label{tras}
\end{eqnarray}
The last equality  defines the adiabatic Hamiltonian  $ H_{WA}$, in which the 
phase $\Phi $ has disappeared (but the explicit time dependence has not !).

By acting  with   $  e^{iS(t) } $ we  relate  the
 Green's functions for the electrons in the lead to the 
adiabatic ones,  $G_{A,ij} $ ($ i,j = L,R $),   corresponding 
to $H_{WA}$, by means of  appropriate phase factors. In particular,
if $t>t' $ and $x,x' \in  ({-L/2}, {L/2} )$ we have:
\begin{eqnarray}
G_{LL}  ( x ,t; x^{'}, t' ) = 
G_{A,LL}  ( x, t; x^{'},  t' ) e^{  i  \int _{t'}^t  \dot{\Phi} (\tau ) d\tau  },  
\nonumber\\
G_{RR}  ( x ,t; x^{'}, t' ) = 
G_{A,RR}  ( x ,t;  x^{'} , t' ) e^{  - i  \int _{t'}^t 
 \dot{\Phi} (\tau ) d\tau  } .
\label{duodeca}
\end{eqnarray}
\noindent
 Eq.s(\ref{tras},\ref{duodeca})  are the third and final remarkable result 
 of this  work. 

Eq.(\ref{tras}) defines the adiabatic Hamiltonian corresponding to our
 simplified model while  Eq.(\ref{duodeca})  relates  the electron
propagators in the interferometer arm to the corresponding one of the
 adiabatic Hamiltonian.

Because of the phase factor  appearing  in the forward propagators of 
Eq.(\ref{duodeca}),  one can infer that  the  
total dc conductance  across the ring, ${\cal G}$, picks up an extra 
interference contribution:
\beq
{\cal G } ( \Phi ) = {\cal G } ( \Phi = 0 ) \times \: 
 \frac{1}{2}  \overline{ \left [ 1 + \cos\left  (\Phi (t+ T) -\Phi (t) 
\right  ) \right ]} \:\:\:\: .
\label{bigass}
\eneq
\noindent
when the dot is driven through a periodic sequence
of adiabatic cycles, with  time period $T \sim L/v_F $  and $b > 0 $.
In Eq.(\ref{bigass}), the overline denotes time 
averaging over a period.

We leave a full derivation of the conductance and its numerical evaluation 
for a forthcoming paper. The  conductance involves the 
 transmission  across   opposite sides
of the contact, between   points at a distance $\sim L $. 
On the other hand   the interference term is not washed out 
by time averaging only if  the traversal time of the electrons propagating 
ballistically is $ \sim T $. This is 
the reason why  one  should have    $T \sim L/v_F $ if  the interference
 has to be constructive. 

The way in which the extra phase $\Phi (t,t')$ appears in
  Eq.(\ref{duodeca}) is a consequence of the local approximation 
made  in order to relate  the  model Hamiltonian derived in Appendix B
with  the toy Hamiltonian of Eq.(\ref{tham}). Actually, in the 
computation  of
Appendix B, space locality had to be relaxed, in order to obtain an 
effective phase transfer from the dot to the wire (this requires the coupling
to take place over a finite length $L$). Time locality, instead, is
much
more meaningful an approximation, since  terms which are nonlocal in time 
may be regarded  as higher time derivatives of the phase and, therefore, 
 are  negligible, due to the hypotesis  of adiabaticity. 
This guarantees that it is always  possible to write down an effective phase 
sensitive Hamiltonian for the capacitive coupling.
 
To conclude this Section,  we have suggested that one 
has a straightforward way to detect  whether or not a Berry phase arises at 
the dot,  independently of the exact value of $\Phi (t) $, and of the 
detailed functional dependence of the electron propagators on $\Phi$. 
Indeed, one may first tune $b<0$, correspondingly measuring 
 ${\cal G} ( \Phi = 0 )$. Then, moving to a working point at $b>0$,
by keeping the working  condition the most possible unchanged,  one may 
measure  ${\cal G } ( \Phi \neq 0  )$ and, eventually, compare the two 
values of the conductance. 

\section{Conclusions}

In conclusion, we have shown that  adiabatically  operating on
three levels of an isolated  quantum dot provides quite a wider 
range of possibilities
than operating on two levels only. In particular, a nonzero Berry
phase $\Gamma$ appears, or not, according to whether $B>B^*$, or $B<B^*$.

It has already been noticed that spin-orbit interaction may generate
a Berry phase \cite{aronov}.
Indeed, electrons moving along a ring in an applied electric
field ${\cal E}$ feel, in their reference frame,
an effective magnetic field, which
gives raise to spin precession.  In our case, the static
$B$ field applied on the dot provides a cyclotron motion, by changing the 
direction of the effective magnetic field along the electronic trajectories.

Our first important result is  that there is no obvious  relation between 
the Berry  phase  arising at  the isolated dot, $\Gamma$, and
the ``dynamical'' phase $\Phi$, appearing in the electron propagator. 
  This is quite a crucial 
point of our analysis. Indeed, while $\Gamma$ is the geometrical phase
acquired by the total wavefunction of dot electrons, $\Phi$ is not a purely
geometrical effect, since it includes dynamical features, that cannot be 
disentangled from the geometrical ones. As shown in Appendix A, this is
a consequence of nonlocality of quantum electron propagation.  As a matter 
of fact, the phase $\Phi$  vanishes, if $\Gamma$ does so.  

In Section IV we  describe a way of detecting the  extra phase arising at
 the dot  by means of a  double path  interferometer.
The setting we propose in Fig.(\ref{fig1}a) is quite peculiar, in that it
works as a Bohm-Aharonov interferometer, although with no oscillating
magnetic flux threading the ring. Instead, our interferometer is controlled by
 voltage tuning at  the quantum dot, which is  located outside of the ring.
 
In Appendix B, we show that an effective phase transfer mechanism needs
a capacitive coupling extended in space.
A local approximation on the  model Hamiltonian derived in 
Appendix  B  leads us to the toy Hamiltonian $H_W^{eff}$ 
 (Eq.(\ref{tham})),  which allows for  
a simple description of the   transferral of the phase  to conduction 
electrons.
Despite its apparent simplicity, the potential in Eq.(\ref{pot}) 
is able to capture the relevant physics arising when there is a nonzero Berry
phase at the dot, and to derive the consequences for the dc conduction 
properties of the ring in quite a straightforward way. In particular, 
because of quantum interference, phase sensitive terms in the electron
propagator do not disappear, as one takes  the local limit.

From the derivation in Appendix B, it is clear that, in
order to transfer the adiabatic phase from the dot electrons to the lead
electrons, the following inequality has to be fulfilled

\[
\frac{\hbar}{E_2 - E_3} <  T \sim \frac{L}{v_F} \;\;\; .
\]
\noindent

The former inequality is the adiabaticity condition: $E_2 - E_3$ is the
separation between the dot levels, that is ${\cal O} ( |g|)$, for $B>B^*$.
It means that cycling must not cause crossings between dot levels. On the
other hand, the propagator for lead electrons takes essential contributions 
nonlocal in space, due to electron self-interference. In order for such an
interference to carry informations about  the total Berry phase at the dot, 
the electron wavefunction must interfere with its opposite chirality 
component, ``coming'' from a distance $L \sim v_F T$. 

A point of view  alternative   to ours is found  in Ref.\cite{gaitan},
where the conduction electrons
are  thought of as  an environment for the dot, in the presence of a Berry 
phase. The bath degrees of 
freedom are integrated out, to provide the adiabatic variable with 
a stochastic  ``force''. In our scheme, instead, we keep the environment 
degrees of freedom (conduction electrons) and calculate the effects of the 
Berry phase on their dynamics.

A device  different from ours has been considered in Ref.\cite{aleiner},
 where a 
QD is embedded in one arm of the interferometer. Usually, such a setup does 
not break phase coherence \cite{ringarm}. In Ref.\cite{aleiner}, however,
  a quantum point 
contact (QPC) is facing the dot, to  detect single-charge tunneling
across the dot as a change in its transmission. 
The detection provides a sudden perturbation and an 
orthogonality  catastrophe in the dot arm, which affects Aharonov-Bohm 
interference in the ring. This is exactly the limit
opposite to ours, since we manipulate the phase of the conducting electrons
by adiabatically operating on a quantum dot at fixed charge.
Moreover, while the dot-QPC coupling of Ref.\cite{aleiner} is point-like, 
in our case, it is extended over a length $L$.

Finally it is important to remark that  we  have 
excluded any pumping effect in the wire   \cite{pumping}.

\vspace*{1cm}

We gratefully acknowledge fruitful discussions with I. Aleiner, B. Jouault,
L. Kouwenhoven, G. Marmo and G. Morandi. 
Work partially supported by contract FMRX-CT98-0180.

\appendix

\section{Adiabatic evolution of the Green's functions of the isolated dot.}

In Section IV we propose to detect the Berry phase of the dot
 by electrostatically 
weakly  coupling  the dot to a conducting wire located near by. In
Appendix B we will derive the effective Hamiltonian for conduction electrons
arising from such a coupling, after integrating out dot's coordinates. 
In particular, the additional couplings involve the  Green's functions for the 
charge density operator of the isolated dot, $\hat q(\underline{r}_b )$, like,
for instance, the time-ordered Green's function:

\beq
G_q ( t , \underline{r}_b ; t^{'} , \underline{r}_b^{'} ) 
= - i  {\rm Tr} \left [ \hat{\rho}  {\cal T}  \hat{q} ( t , \underline{r}_b ) 
\hat{q} ( t^{'}  , \underline{r}_b^{'}  ) \right  ]
\label{app2}
\eneq
\noindent
where $\hat{\rho}$ is the density  matrix for the isolated dot and
${\cal T}$ is the time ordering operator. 

In this Appendix we will derive the Green's functions for the isolated dot 
undergoing adiabatic evolution  of Eq.(\ref{app2}).
 In particular, we will show how $G_q$
is related to its adiabatic counterpart, $G_{q,A}$, constructed from 
the adiabatic Hamiltonian $H_A$. 

Since the  coordinate $ \underline{r}_b $,  confined to the boundary of the 
dot, is immaterial in the following, we just drop it  henceforth. 

The adiabatic Hamiltonian 
$H_A (s)$  has been  introduced in Section III. Its  eigenstates, 
$| e_m ( s ) \rangle$, solve, with eigenvalues $E_m (s)$,  the 
stationary Schr\"odinger equation for $H_D (t)$ (shortly denoted by $H (t)$ in
the following) with  $t$ fixed and equal to $Ts$:
\beq
 H_A (s) | e_m(s) \rangle  = E_m (s) |e_m(s) \rangle  . 
\eneq
\noindent

In Section III we have shown that, when the isolated dot is driven along
a closed path in parameter space, $\gamma = \vec{\lambda} ( s ) \; ( 0 \leq s 
\leq 1)$, its state $ |F ( t = 0 ) \rangle$ evolves into 
$ | F ( T ) \rangle$, by picking a Berry phase $\Gamma$, that
is, $ | F ( T ) \rangle = e^{ i \Gamma} |F ( 0 ) \rangle$. $\Gamma$
is given by (see Eq.(\ref{piuone})):

\beq
\Gamma = i  \oint_\gamma d \lambda^a   \biggl \langle e_0 ,
\vec\lambda   \biggl| 
\frac{ \partial}{\partial \lambda^a } \biggr|  e_0 ,\vec\lambda 
 \biggr\rangle \;\;\; .
\label{ap1}
\eneq
\noindent
In Eq.(\ref{ap1}),   $| e_0 , \vec{\lambda}  \rangle$ coincides with $|e_3,
\vec{\lambda} \rangle $ 
of Section III. The sum over repeated indices $a$ is understood.

In order to relate the full Hamiltonian $H (t)$ to  $H_A$, let us consider 
the projector $P_m (s) = 
| e_m (s ) \rangle \langle e_m (s)|$. Since during the adiabatic evolution,
level crossings are forbidden, we have  $ [ H (T s ) , P_m (s) ] = 0$. 
$P_m$ evolves in time according to:
\beq
P_m ( s ) = U_A^\dagger ( s ) P_m ( 0 ) U_A ( s ) \;\;\; \forall s
\label{eqtwo}
\eneq
\noindent
where $U_A ( s ) = U_A ( s , 0 ) = 
{\cal T}  \exp [ - i T \int_0^s d \sigma H_A ( \sigma )]$.

By following the derivation of Ref.\cite{avron0}, it is possible to prove that
$H_A ( s )$ is related to $ H ( s T )$ according to:

\beq
 H_A ( s )  =  H ( T s )  + \frac{i}{T} \sum_m [ \dot{P}_m ( s ) , P_m ( s ) ]
\label{ap2}
\eneq
\noindent
where $\dot{P}_m ( s ) = \frac{ d}{d s } P_m ( s )$.

In order to compute the time derivatives of $P_m (s)$, let us employ the
integral representation of the projector in terms of the analytic extension
of the Green's operator to complex energies, $G_A ( z ;s ) = 
( z - H_A ( s ) )^{-1} $:  
\beq
P_m ( s ) = \oint_{r_m} \frac{ d z }{ 2 \pi i } G_A ( z ; s ) .
\label{ap4}
\eneq
\noindent
where $r_m$ is a closed path encircling $E_m ( s )$. 

From Eq.(\ref{ap4}), it is straightforward to prove that:

\beq
\dot{P}_m ( s ) = \oint_{r_m}  \frac{ d z }{ 2 \pi i } G_A ( z ; s )
\dot{\lambda}^a ( s ) \frac{ \partial H}{ \partial \lambda^a }
G_A ( z ; s ) \;\; , 
\label{ap5}
\eneq
\noindent

Therefore, we obtain the following representation for $[ \dot{P}_m , 
P_m]$:

\begin{eqnarray}
[ \dot{P}_m , P_m ] ( s )  &=& \sum_{n ( n \neq m )} 
\biggl\{ | e_n ( s )  \rangle 
\frac{ \langle e_n  ( s ) | \dot{\lambda}^a \frac{ \partial H}{ 
\partial \lambda^a } | e_m ( s ) \rangle}{ \omega _{mn} ( s ) } 
\langle e_m ( s ) |  \nonumber \\
&-&  | e_m ( s )  \rangle 
\frac{ \langle e_m  ( s ) | \dot{\lambda}^a \frac{ \partial H}{ 
\partial \lambda^a } | e_n ( s ) \rangle}{\omega _{mn} ( s ) } 
\langle e_n ( s ) | \biggr\}
\label{addapp1}
\end{eqnarray}
\noindent
 where $ \omega _{mn} ( s ) =  E_m ( s ) -E_n ( s )$.
In the following part of this Appendix, we will use the short-hand
notation $ | m_s \rangle$, rather than $| e_m (s )\rangle$.

Eq.(\ref{addapp1}) shows, in particular, that $[ \dot{P}_m , P_m]$ is
a fully off-diagonal operator, as it must be. 
In the following, we will use its off-diagonal matrix elements:

\begin{eqnarray}
\langle m | \left [ \dot {P}(s) , P(s) \right ] |n\rangle =
 - 2   \frac{\dot\lambda _s^a }{\omega _{mn}}
\langle m |  \frac{\partial  H}{\partial \lambda ^a  } | n \rangle  \nonumber\\
\hspace{1cm}    (  m \neq n ).
\label{ris}
\end{eqnarray}

Another important operator is $\Omega ( s ) = U_A^\dagger ( s ) U ( T s )$. It
may be shown  that $\Omega ( s ) = 1 + {\cal O} (1 / T )$, but its 
 derivative  is  $ {\cal O } ( 1 )$:
\begin{eqnarray}
\dot{ \Omega} ( s ) = - U_A^\dagger ( s ) \sum_m [ \dot{ P}_m (s ) , 
P_m ( s ) ] U_A ( s ) \Omega ( s )
\label{ap3}
\end{eqnarray}
\noindent

The time-ordered Green's function for the observable  $ \hat{q}$  
may be written as:

\begin{eqnarray}
G_q ( t , t^{'} ) = - i \theta ( t - t^{'} ) {\rm Tr} \left [ \hat{\rho} 
\Omega^\dagger ( s ) \hat q_A ( s ) \Omega (s) \Omega^\dagger ( s^{'} ) 
\hat q_A ( s^{' } ) \Omega ( s^{'} ) \right ]
\nonumber\\
- i \theta ( t^{'} - t  ) {\rm Tr} \left [ \hat{\rho} 
 \Omega^\dagger ( s^{'} ) 
\hat q_A ( s^{' } ) \Omega ( s^{'} ) 
\Omega^\dagger ( s ) \hat q_A ( s ) \Omega (s) \right ]
\nonumber
\end{eqnarray}
\noindent
where we have written down the time ordering prescription explicitly.
The adiabatic operator is defined as $\hat{q}_A  ( s ) = U_A ( s ) \hat{q} 
U_A^\dagger ( s ) $.

To ${\cal O} ( \frac{1}{T} )$, the equation of motion for $G_q(t,t') $  
will contain terms depending on $\dot{\Omega} (s) $: 

\begin{eqnarray}
i \partial_t G_q (t,t' )  = 
   {\rm Tr} \left [  {\cal{T}}  
[ H_A ( s ) , \hat{q}_A ( s ) ]  \hat{q}_A (s^{'} ) \ro  \right ]
\nonumber\\ 
 +  \delta ( t - t^{'} ) 
 +   {\rm Tr} \left [  {\cal{T}} \ro  \Omega ( s')  
 \dot{\Omega} ^\dagger (s)  q_A(s)  \Omega ( s)
  \Omega ^\dagger (s')  q_A(s')  \right ] 
\nonumber\\
 +     {\rm Tr} \left [  {\cal{T}}
  \ro  \Omega ( s') 
 \Omega ^\dagger (s)  q_A(s)  \dot{\Omega }( s)
  \Omega ^\dagger (s')  q_A(s')  \right ]\nonumber\\
\label{appad1}
\end{eqnarray}
\noindent
The statistical weight is  $ \rho_{\Omega}(s') =  \Omega ( s') \rho 
 \Omega^\dagger  ( s')$. To lowest order in $1/T$ it coincides with 
the equilibrium  density matrix $\rho $ at $t =0$.

Finite-temperature Wick theorem allows us to write down the second and third 
line at the  r.h.s. of Eq.(\ref{appad1}) as:
\begin{eqnarray}
i{\rm Tr} \left [ {\cal{T}} \Omega (s') \dot{\Omega}^\dagger ( s ) \ro \right ]
G_{q , A} ( s , s^{'} ) \nonumber\\
+ i  {\rm Tr} \left [ {\cal{T}} \dot{\Omega} ( s )\Omega ^\dagger(s')
 \ro \right ]
\tilde{G}_{q , A} ( s , s^{'} ) \;\; ,
\label{appaddd2}
\end{eqnarray}
\noindent
where we have introduced the time-ordered Green's function of 
$\hat{q}_A ( s )$, $G_{q,A}$, and the anti time-ordered Green's function, 
$\tilde{G}_{q,A}$.
This amounts to evaluate the average along a Keldysh contour \cite{rammer}.
$G_{q,A}$ and   $\tilde{G}_{q,A}$  are explicitly given by:
\begin{eqnarray}
G_{q,A} ( s , s^{'} ) = - i  {\rm Tr} \left [
{\cal T}  \hat{\rho} (T)  \hat{q}_A ( s ) \hat{q}_A ( s^{'}  )  \right ] 
\nonumber \\
\tilde{G}_{q,A} ( s , s^{'} ) = - i  {\rm Tr} \left [
\tilde{\cal T}  \hat{\rho} (0) \hat{q}_A ( s ) \hat{q}_A ( s^{'}  )  \right ] 
\end{eqnarray}
\noindent
where  $\tilde{\cal T}$  is the anti-time ordering operator. 
 
In order to close the set of equations of motion, we need the 
analog of Eq.(\ref{appad1}) for $\tilde{G}_{q  }$:
\begin{eqnarray}
i \partial_t \tilde{G}_q (t , t' )  = 
   {\rm Tr} \left [ \tilde{\cal{T}}  \hat \rho (0)  
[ H_A ( s ) , \hat{q}_A ( s ) ]  \hat{q}_A (s^{'} ) \right ]
\nonumber\\
 - \delta ( t - t^{'} ) 
 + i {\rm Tr} \left [\tilde {\cal{T}} \dot{\tilde{\Omega}} ( s )
\tilde{\Omega} ^\dagger (s') \rot \right ] 
\tilde{G}_{q,A} ( s , s^{'} )
\nonumber\\
 + i {\rm Tr} \left [\tilde{\cal{T}}
\rot \tilde{\Omega} ( s' ) 
\dot{\tilde{\Omega}}^\dagger ( s ) \right ] G_{q,A} ( s , s^{'} ) 
\nonumber
\end{eqnarray}
\noindent
where $ \tilde{\Omega }(s)  = U_A ( s , 0 ) U ( 0 , s)$ and
 $ \rot =   \tilde{\Omega } ( s') \rho 
  \tilde{\Omega }^\dagger  ( s')$.
Let us define :
\beq
\partial _s\Phi (s,s') \equiv 
i {\rm Tr} \left [ {\cal{T}}\:
\dot{\Omega} ( s ) \Omega ^\dagger (s') \: 
\rho_{\Omega}(s') \right ]
\label{fas}
\eneq  
It is immediately seen  that
\beq
\partial _s\tilde{\Phi} (s,s') \equiv 
i {\rm Tr} \left [\tilde {\cal{T}}\: \dot{\tilde \Omega} ( s )
 \tilde{\Omega} ^\dagger (s') \: \rot \right ]
= \partial _s\Phi^\dagger (s,s') \:  . 
\label{ilde}
\eneq 
With the definitions of Eq.s (\ref{fas},\ref{ilde}), the set of equations of 
motion for the $G_q$'s now reads:
\begin{eqnarray}
i \partial _t G ( t , t^{'} ) = \delta (t-t') 
+  H_A(s) G_A ( s , s^{'} ) \nonumber\\
 +\left [ \partial _s \tilde{\Phi}  (s,s') \:  
    G_A ( s , s^{'} )    
 - \partial _s {\Phi}  (s,s') \: \tilde{G}_A (s , s' )\right ] \;\; ;
\nonumber\\
i \partial _t \tilde{G} ( t , t^{'} ) = - \delta (t-t') 
+  H_A(s)  \tilde{G}_A ( s , s^{'} )\nonumber\\
 +\left [ \partial _s { \Phi}  (s,s') \:  {G}_A ( s , s^{'} )
 - \partial _s \tilde { \Phi}  (s,s') \:  G_A (s , s' ) \right ]
\label{system2}
\end{eqnarray}

We now use the definitions:
\begin{eqnarray}
G^>(t,t') - \tilde{G}(t,t') \equiv G^{(R)}(t,t')\nonumber\\ 
G^<(t,t') - \tilde{G}(t,t') \equiv G^{(A)}(t,t')\nonumber\\
G(t,t') + \tilde{G}(t,t') \equiv G^{(K)}(t,t')\equiv G^>(t,t') +
 G^<(t,t').  \nonumber
\end{eqnarray}

Subtracting, for $s>s'$, the two Eq.s(\ref{system2}), we obtain:
\begin{eqnarray}
i \partial _t G^{(R)} ( t , t^{'} ) = \delta (t-t') 
+  H_A(s)  G^{(R)}_A ( s , s^{'} )  \label{gr}\\
 +\left ( \partial _s {\tilde \Phi}  (s,s') -  \partial _s { \Phi}  (s,s') 
\right )  G^{(K)}_A ( s , s^{'} ) \nonumber
\end{eqnarray}
We proceed in the same way for the advanced Green's function $ G^{(A)}$,
 in the case $s<s'$.

The sum of  Eq.s (\ref{system2}) gives:
\begin{eqnarray}
i \partial _t G^{(K)} ( t , t^{'} ) = 
  H_A(s)  G^{(K)}_A ( s , s^{'} ) \nonumber\\
 + \left ( \partial _s { \Phi}  (s,s') + \partial _s {\tilde \Phi}  (s,s') 
\right )  G^{R,A}_A ( s , s^{'} ) \label{gk}
\end{eqnarray}
where the retarded or advanced Green's function appears on the 
r.h.s.   depending on whether $s>s'$, or $s<s'$.

The derivatives w.r.to $t'$ can be  worked out in the same way. By changing 
$ t$ and $ t' $, $\Omega ^\dagger $ and $\Omega $ are exchanged. This 
implies a minus sign:
 \begin{eqnarray}
i \partial _{t'} G^{(R)} ( t , t^{'} ) = \delta (t-t') 
+  G^{(R)}_A ( s , s^{'} ) H_A(s') \nonumber\\
 - \left [ \partial _{s'}
 {\tilde \Phi}  (s,s') - \partial _{s'} { \Phi}  (s,s') 
\right ]  G^{(K)}_A ( s , s^{'} ) 
\end{eqnarray}

 Let us  define:
\beq
\partial _s\varphi _\pm  ( s,s' )\equiv 
  \partial _s \tilde { \Phi}  (s,s') \pm  \partial _s { \Phi}  (s,s')\:\: . 
\label{varf}
\eneq

From now on we choose $s>s' $.  Using the cyclic invariance of the trace 
we see that: 
\begin{eqnarray}
 \partial _s { \Phi}  (s,s') =  - i
 \biggl\langle F_{0 s'}\biggl | U_A(s',s) \:   [  \dot{P}(s),P(s) ]\:
  \biggr | F_{s 0} \biggr\rangle \nonumber\\
 \partial _s \tilde { \Phi}  (s,s') =
i \biggl\langle F_{1 s'}\biggl | U_A(s',s) \:   [  \dot{P}(s),P(s) ]\:
  \biggr | F_{s 1} \biggr\rangle  \:\:\:  , 
\nonumber
\end{eqnarray}
where  $  | F_{s 0} \rangle = U(s,0 )   | o \rangle $ 
( $  | F_{s 1} \rangle  =  U(s,1 )   | o \rangle $),
and we assume we are working at zero temperature.
 
 We now calculate   $  | F_{s 0} \rangle $,
 $  | F_{s 1} \rangle $ .

\vspace*{0.3cm}

They solve the   Schr\"odinger equation:
\beq
i \frac{d |F_s \rangle }{ds} = T H |F_s \rangle
\label{schro}
\eneq 
with different initial conditions.
We expand   the ground state  on the adiabatic basis set:
\begin{eqnarray}
 |F_{s 1} \rangle = U(s,1)|F_1 \rangle =  e^{i\Gamma } \biggl\{
 e^{-iT\int _1 ^{s} E_o} |o_s \rangle \nonumber\\
 + \sum _{l\neq o} a_{lo}(s)
 e^{- iT\int_1^s  E_l  }\:|l_s \rangle  \biggr\}
\nonumber \\
 |F_{s 0} \rangle = U(s,0)|F_0 \rangle =   \biggl\{
 e^{-iT\int _0 ^{s} E_o} |o_s \rangle  \nonumber\\
 + \sum _{l\neq o} b_{lo}(s)
 e^{- iT\int_0^s  E_l  }\:|l_s  \rangle  \biggr\}
\nonumber
\end{eqnarray}
where $\Gamma $ is the Berry phase.
By inserting these states  in Eq.(\ref{schro}), we obtain the differential 
equations for the coefficients  $ a_{lo} , \: b_{lo}$.  These are  solved 
 perturbatively,   by taking, to zero order,  $ a_{lo}^0(s)= b_{lo}^0(s)
 = \delta _{lo}$.  Using 
the initial condition  $ a_{lo}(1) = \delta _{lo}$, we find:
\begin{eqnarray}
a_{lo} (s)  = \int _s^1 \dot{\lambda}^a_{s'} \: ds'
 \frac{e^{iT\int _{s'}^1 \omega _{ol} } }{ \omega _{lo} } \langle l_{s'}|  
 \frac{\partial  H}{\partial \lambda ^a }|o_{s'} \rangle  \:\:\:\:\:  l\neq o 
\label{alo}
\end{eqnarray}
in the first case.  In the same way, with $ b_{lo}(0) = \delta _{lo}$, 
we get:
\begin{eqnarray}
b_{lo} (s)  =- \int _0^s \dot{\lambda}^a_{s'} \: ds'
 \frac{e^{iT\int _0^{s'} \omega _{lo} } }{ \omega _{lo} } \langle l_{s'}|  
 \frac{\partial  H}{\partial \lambda ^a }|o_{s'} \rangle  \:\:\:\:\:  l\neq o 
\label{bilo}
\end{eqnarray}
in the second case.

By inserting  Eq.s (\ref{ris},\ref{alo},\ref{bilo})
 in $\varphi _{\pm} ( s,s' )$, given by 
 Eq.(\ref{varf}), to lowest order we obtain the following four terms:

\begin{eqnarray}
\partial _s \varphi _+ =  -i(\int _{s'}^1 + \int _0^{s'} )\: Z^*_s +
 i(\int _{s}^1 + \int _0^{s} ) \: Z_s  \nonumber\\
\equiv  -2\: \Im m\int _0^1 \: d\tau \: Z_{s}(\tau  )
  \equiv \dot {\Phi} _o
\nonumber\\
\partial _s \varphi _- =  -i(\int _{s'}^1 - \int _0^{s'} ) \: Z^*_s 
 + i(\int _{s}^1 -\int _0^{s} ) \: Z_s 
\end{eqnarray}
with
\begin{eqnarray}
   Z_{s} (\tau ) =  2\dot\lambda _s^a \: \dot{\lambda ^b}_{\tau} \: 
  \sum _{m \neq o } \: 
 \langle o_{\tau}|  
 \frac{\partial  H}{\partial \lambda ^b  }|m_{\tau } \rangle 
\nonumber\\
 \times \frac{e^{-iT\int _s^{\tau }d\tau \:
 \omega _{mo} } }{\left ( \omega _{mo} \right )^2 }
\langle  m_s |  \frac{\partial  H}{\partial \lambda ^a }|o_s \rangle 
\end{eqnarray}
$\dot {\Phi} _o $ is independent of $s, s'$  because of our  assumption  that  
$\Omega (s)\Omega ^\dagger (s') \approx 1 $. 

The integral  of $  \dot {\Phi}_o $ w.r.to $s$  can be rewritten as: 
\begin{eqnarray}
\Phi _o  = 2 \:\Im m \int_0^1 ds \: \int _0^1 ds'' \:   \;
 (  \dot{ \lambda}^a (s)  \dot{ \lambda}^b (s'') \; )
  \: \sum_{ m \neq 0 } \label{faso}\\
\langle  e_0 ( s^{''} ) |\frac{ \partial H}{ \partial \lambda^a}
 | e_m ( s^{''} ) \rangle  \;
\frac{ e^{ i T \int_s^{s^{''}}\!\! d\tau \; \omega_{0m}
 ( \tau )}}{\left ( \omega _{mo} \right )^2 } \:
 \langle e_m ( s ) | \frac{ \partial H}{\partial \lambda ^b} |
  e_0 ( s ) \rangle \;\; ,
\nonumber
\end{eqnarray}
\noindent
where  the symbols are defined in Eq.(\ref{eqtenbis}).
We see that $\Phi _o $ resembles the
result of Eq.(\ref{eqtenbis}), but it takes the form of a non local Berry phase.

On the other hand $\partial _s \varphi _- \to \dot {\Phi} _o $ only in the 
limit $  s' \to 0^- ,  s\to 0^+ $:
\beq
\partial _s \varphi _- ( s\to 0^+ , s' \to  0^- )  \to    \dot {\Phi} _o
\eneq
 In the limit $  s' \to 0^- , 
 s\to 0^+ $,  the motion equations become:
\begin{eqnarray}
{\cal{L}}(  G^> - G^< ) =  \dot {\Phi} _o \: ( G^> + G^< )  \nonumber\\
{\cal{L}}(  G^> + G^< ) =  \dot {\Phi} _o \:  ( G^> - G^< )  \nonumber\\
\end{eqnarray}
where we have used the  
short hand notation ${\cal{L}}$, to denote the differential operator of Eq.s
(\ref{system2}). This implies that:
\begin{eqnarray}
{\cal{L}}  G^>  =   \dot {\Phi} _o  G^> \:\:,  \hspace{0.5cm}
{\cal{L}}  G^<  =  -  \dot {\Phi} _o    G^<  \:\: , \nonumber
\end{eqnarray}
or 
\begin{eqnarray}
G^{(>,<)} ( t , t^{'} ) = G_{A}^{(>,<)} ( s , s^{'} ) 
 e^{ \pm i (t-t' )\cdot  \dot {\Phi} _o  }\nonumber\\
  \hspace{1cm} (s\sim  s' \sim 0 ,
  \: s>s' )
\label{gminmag}
\end{eqnarray}
 Hence, $ \dot {\Phi} _o $ appears in the phase factor relating 
the dot's Green's functions to the adiabatic ones,
 at least at small time intervals. 
When $ s-s' $ increases, $ G^{(>)}$ and $G^{(<)}$  are mixed together and 
the solution of the system of Eq.(\ref{gr}, \ref{gk}) is not straightforward. 

\section{The effective Hamiltonian for the  electrons  in the lead}

In this Appendix  we calculate the
effective propagator for electrons in the arm facing the dot,  given by
Eq.(\ref{Z}). By  integrating out the dot degrees of freedom, we 
construct an effective action. This action  describes
an interaction among lead  electrons, mediated by the quantum dot. Here,
we will show that it may be approximated by a mean field self-consistent 
effective potential, acting on the lead electrons,  and we derive in
detail  the various contributions. This shows that the model Hamiltonian
in Eq.(\ref{tham}) does, in fact, entail the relevant features of
the $\Phi$-dependent term in the effective Hamiltonian.

As we have outlined in Section IV, we perform the integration over dot's 
variables within Feynman's action formalism. The following 
four-Fermion effective action arises:

\begin{eqnarray}
 S_{int} =
 -\met   \int dt \:dt' G_q (t , t^{'} )  \int dx dx'\times
 \nonumber \\
 f(x) f(x')  
 [   \Psi ^\dagger (x,t)  \Psi (x,t)
  \Psi ^\dagger (x',t')   \Psi (x',t') ] \;\;\; .
\label{cumulb}
\end{eqnarray}

$G_q ( t , t^{'} )$  appearing in  Eq.(\ref{cumulb})
 is  the time ordered Green's function for the isolated dot, that we have 
derived in Appendix A.  

Since we are approximating the wire as a one-dimensional conductor, we may
write  the Fermionic field as a linear combination of the left-handed and
of the right-handed fields, introduced in Eq.(\ref{lr}):

\beq
\Psi (x) = e^{ - i q_F x } \psi_L ( x ) + e^{ i q_F x } \psi_R ( x )
\label{lrb}
\eneq
\noindent

We decouple the  four-Fermion term in Eq.(\ref{cumulb}) by means of 
a Hubbard-Stratonovitch transformation. Then, we perform a saddle-point
approximation, where we express the auxiliary Bosonic
fields in terms of self-consistent pairwise
averages of  the Fermionic fields. In particular, in order to make the
derivation the simplest is possible, we take the  Green's functions  
$  \langle \langle  \psi_i ( x , t ) \psi_j ^\dagger ( x', t' )\rangle 
\rangle  , ( i,j = L,R) $ to be equal to the ones  for electrons scattered by 
a point-like scattering center at the origin. 

The various kernel appearing in the effective action are nonlocal, 
both in time and space. In order to resort to an Hamiltonian that is local
in time, we single out the most relevant trajectories of conduction electrons,
by taking the local-time limit (``local-time approximation''). Following such a 
procedure gives raise to  an  effective $L-L$ and  $R-R$ ``forward '' 
scattering contribution,  as well as  to  $L-R$ and $R-L$ ``backscattering'' 
terms. We expand for $ x \sim x' $ close to the origin where the overlap 
$ f(x) f(x') $ is maximum.  In the local approximation we assume that 
Eq. (\ref{gminmag})  can be extensively used. 
Let us now show the details of the derivation:

{\sl term a) }:
A first contribution to the $L-L$ scattering is given by the following 
decoupling pattern:

\begin{eqnarray}
S_{LL}^{(a)} = - \frac{1}{2} \int d t \; d t^{'} \int d x \; d x^{'} \;
f ( x ) f (x^{'} )  G_q ( t , t^{'} )  \times \nonumber
\\
\biggl\{ \psi_L^\dagger ( x , t ) \psi_L ( x^{'} , t^{'} ) [ \; 
\langle \langle \psi_L  ( x , t ) \psi_L^\dagger ( x^{'} , t^{'} ) \rangle
\rangle + 
\nonumber \\  
\langle \langle 
\psi_R  ( x , t ) \psi_R^\dagger ( x^{'} , t^{'} )\rangle \rangle 
 e^{  2 i q_F (x-x^{'} ) }  \; ]  
\nonumber 
\\
- \psi_L^\dagger ( x^{'} , t^{'} )  \psi_L ( x , t ) [ \; 
\langle \langle  \psi_L^\dagger ( x , t )  \psi_L ( x^{'} , t^{'} ) 
\rangle \rangle +
\nonumber \\
\langle \langle  \psi_R^\dagger ( x , t )  
\psi_R ( x^{'} , t^{'} )\rangle \rangle
 e^{ -2 i q_F (x - x^{'}) }  \; ] \biggr \} \:\: . 
\label{terab}
\end{eqnarray}
\noindent
Here we take  $ \langle \langle \psi_{L(R)} ( x , t )\psi_{L(R)}^\dagger
 ( x^{'} , t^{'} ) 
\rangle \rangle  =   -i/ [ x -x' \mp v_F (t -t' ) + i 0^+ ]^{-1} $ ( where 
the upper (lower ) sign is for $L (R)$), and
 $ \langle \langle  \psi_{L(R)}^\dagger ( x , t ) 
 \psi_{L(R)} ( x^{'} , t^{'} )
\rangle \rangle   = i [ x -x' \mp v_F (t -t' ) + i 0^+ ]^{-1} $.
We estimate $G_q(t,t') $ in the local-time approximation, 
 at    $ | t-t' |  \approx  | x-x' | /v_F $.
Only the principal part  ${\cal P}$  of the  integral 
 survives  in Eq.(\ref{terab}),
giving:
\begin{eqnarray}
S_{LL}^{(a)} =  \int d t \; \int d x \; d x^{'} 
f(x) \; f (x^{'} ) \psi_L^\dagger ( x , t ) \psi_L ( x^{'} , t )
\nonumber \\
\times e^{ i q_F ( x - x^{'} )} {\cal P} \biggl[ \frac{ \sin  q_F ( x -
x^{'} )}{ x - x^{'} } \biggr] \nonumber \\
 \times \biggl[ G_q ( t + \frac{x - x^{'} }{ v_F} , 
 t ) + G_q ( t - \frac{x - x^{'} }{ v_F} , t ) \biggr]\:\:  .
\label{loctb}
\end{eqnarray}
\noindent

We further expand the term in Eq.(\ref{loctb}) involving the dot
  Green's functions  about $x-x^{'} \sim 0$, to give the final result:

\begin{eqnarray}
S_{LL}^{(a)} =  2 \int d t \; \int d x \; d x^{'} 
f(x) \; f (x^{'} ) \psi_L^\dagger ( x , t) \psi_L ( x^{'} , t )
\nonumber \\
\times  e^{ i q_F ( x - x^{'} )} {\cal P} \biggl[ \frac{ \sin ( q_F ( x -
x^{'} ) )}{ x - x^{'} } 
\biggr] \Re e \{ G_{q , A}^{\rm Ret} ( t^{+}, t ) \} 
\label{finalr1}
\end{eqnarray}
\noindent
A similar term  arises if  the same decoupling pattern is used for the
right-handed field. Putting both together, we obtain 
an explicitly time-dependent Hamiltonian contribution in the form:

\begin{eqnarray}
V_a ( t ) = v_a ( t )  \int d x \; d x^{'} 
f(x) \; f (x^{'} ) {\cal P} \biggl[ 
\frac{ \sin  q_F ( x - x^{'} )  }{ x - x^{'} }  \biggr] 
\nonumber \\
 e^{ i q_F ( x - x^{'}  )}
[ \psi_L^\dagger ( x , t ) \psi_L ( x^{'} , t ) + 
 \psi_R^\dagger ( x , t ) \psi_R ( x^{'} , t )  ]
\label{finalr2}
\end{eqnarray}
\noindent
where $v_a (t) = \Re e \{ G_{q , A}^{\rm Ret} ( t^{+}, t ) \} $.

The contribution above is a purely electrostatic  forward scattering
potential, independent of  $\Phi$. As it is not affected by the adiabatic 
phase, we have dropped it in the discussion of the effective Hamiltonian
of Section IV, Eq.(\ref{tham}).

{\sl term b) }:
A second possible decoupling pattern 
 contributes to an $L-L$ scattering potential, as well:
\begin{eqnarray}
S_{LL}^{(b)} = - \frac{1}{2} \int d t \; d t^{'} \int d x \; d x^{'} \;
f ( x ) f (x^{'} )  G_q ( t , t^{'} )  \times \nonumber
\\
\{ \psi_L^\dagger ( x , t ) \psi_L ( x^{'} , t^{'} ) [ \; 
\langle \langle \psi_R  ( x , t ) \psi_L^\dagger ( x^{'} , t^{'} ) \rangle
\rangle e^{ 2 i q_F x} + 
\nonumber \\  
\langle \langle 
\psi_L  ( x , t ) \psi_R^\dagger ( x^{'} , t^{'} )\rangle \rangle 
 e^{ - 2 i q_F x^{'} }  \; ]  
\nonumber 
\\
- \psi_L^\dagger ( x^{'} , t^{'} )  \psi_L ( x , t ) [ \; 
\langle \langle  \psi_R^\dagger ( x , t )  \psi_L ( x^{'} , t^{'} ) 
\rangle \rangle e^{- 2 i q_F x }
+
\nonumber \\
\langle \langle  \psi_R^\dagger ( x , t )  
\psi_L ( x^{'} , t^{'} )\rangle \rangle
 e^{ 2 i q_F x^{'} }  \; ] \}
\label{termineb}
\end{eqnarray}
\noindent
Here we take  $ \langle \langle \psi_{L(R)} ( x , t )\psi_{R(L)}^\dagger
 ( x^{'} , t^{'} ) 
\rangle \rangle  =   i/ [ x +x' \mp v_F (t -t' ) - i 0^+ ]^{-1} $. 
In the local-time approximation, we consider
only fluctuations around $ | x + x^{'} |  \sim \pm v_F | t - t^{'} |$.
Following the same steps as before we get:
\begin{eqnarray}
S_{LL}^{(b)} = \int d t \; \int d x \; d x^{'} \; f ( x ) 
f ( x^{'} ) \psi_L^\dagger ( x , t ) \psi_L ( x^{'} , t ) 
\nonumber \\
e^{ i q_F ( x - x^{'} ) } {\cal P } \biggl[ \frac{ \sin ( q_F ( x + x^{'} ))}{
x + x^{'} } \biggr] 
\times \nonumber \\
\biggl[ G_q ( t + \frac{ x + x^{'} }{v_F} , t ) 
-  G_q ( t - \frac{ x + x^{'} }{v_F} , t ) \biggr]
\label{eqtermineb2}
\end{eqnarray}
\noindent
Differently from Eq.(\ref{loctb}), Eq.(\ref{eqtermineb2}) contains the
difference between the $G_q$'s. This gives raise to a first-order contribution
in $\dot{\Phi}$, 
when expanding  the last row of Eq.(\ref{eqtermineb2}),
according to Eq.s(\ref{gminmag}). The result is:
\begin{eqnarray}
S_{LL}^{(b)} = \frac{2}{v_F} 
\int d t \Im m \{ G_{q , A}^{\rm Ret} ( t^+ , t) \} \dot{\Phi}  ( t )
\times \nonumber \\
 \int d x \; d x^{'} \; f ( x ) 
f ( x^{'} )  \sin ( q_F ( x + x^{'} )  )  \psi_L^\dagger ( x , t ) 
\psi_L ( x^{'} , t ) 
\label{eqtermine3}
\end{eqnarray}
\noindent

where $  2 \: \Im m \{ G_{q , A}^{\rm Ret} ( t^+ , t )\} /v_F$ is a 
 slowly-varying  function of time.

Since the right-moving fields enter the forward scattering term with the same 
expression, except for the replacement $v_F \rightarrow - v_F$, the total
$\Phi$-dependent forward scattering potential  for the lead electrons will be:

\begin{eqnarray}
\tilde{V}_\Phi ( t ) =   
\dot{\Phi} ( t ) \int d x \; d x^{'} \; f ( x ) f ( x^{'} )
\:\: \frac{2}{v_F}  \: \Im m \{ G_{q , A}^{\rm Ret} ( t^+ , t )\} 
\nonumber\\
\sin ( q_F ( x + x^{'} )) 
[ \psi_L^\dagger ( x ,t ) \psi_L ( x^{'} , t ) -
 \psi_R^\dagger ( x ,t ) \psi_R ( x^{'} , t ) ]
\nonumber
\end{eqnarray}
\noindent
This term  is phase sensitive   and corresponds to 
the local potential $ V_\Phi (t)$ in $H^{eff}_W$  (Eq.(\ref{pot})).  

{\sl term c)}:

Here, we consider the effective action generated by the contraction
pattern:

\begin{eqnarray}
S_{LR}^{(c)} = 
- \frac{1}{2} \int d t \; d t^{'} G_q ( t , t^{'} ) \int d x \; d x^{'} 
\; f ( x ) \; f ( x^{'} ) \nonumber \\
\times \{ \psi_L^\dagger ( x , t ) \psi_R ( x^{'} , t^{'} ) [  \; 
 \langle \langle \psi_L ( x^{'} , t^{'} ) \psi_L^\dagger ( x , t ) 
\rangle \rangle e^{ 2 i q_F  x^{'}}
\nonumber \\
 + \langle \langle \psi_R 
( x^{'} , t^{'} ) \psi_R^\dagger ( x , t )  \rangle \rangle e^{ 2 i q_F x} \; ]
\nonumber \\
- \psi_L^\dagger ( x^{'} , t^{'} ) \psi_R ( x , t ) 
[ \; \langle \langle \psi_L^\dagger ( x , t ) \psi_L ( x^{'} , t^{'} ) \rangle
\rangle e^{ 2 i q_F x } 
\nonumber \\
+ \langle \langle \psi_R^\dagger ( x , t ) 
\psi_R (x^{'} , t^{'} ) \rangle \rangle e^{ 2 i q_F x^{'}} 
\; ] \}
\label{offdiag1}
\end{eqnarray}
\noindent
In the local-time approximation, we get:

\begin{eqnarray}
S_{LR}^{(c)} = \int d t \; \int d x \; d x^{'} \; f ( x ) \; f ( x^{'} ) 
\; e^{ i q_F ( x + x^{'} )}
\nonumber \\
\times  \psi_L^\dagger ( x , t ) \psi_R ( x^{'} , t)
{\cal P} \biggl[ \frac{ \sin ( q_F ( x - x^{'} ) )}{ x - x^{'} } 
\biggr] 
\times \nonumber \\
\biggl[ G_q ( t + \frac{ x - x^{'}}{v_F} , t ) - 
 G_q ( t - \frac{ x - x^{'}}{v_F} , t ) \biggr] 
\label{offidag2}
\end{eqnarray}
\noindent

Expansion of these term gives zero, when $x \rightarrow x^{'}$.

{\sl term d)}: 

The last contraction pattern gives raise to the following term:

\begin{eqnarray} 
S_{LR}^{(d)} = - \frac{1}{2} \int d t \; d t^{'}  G_q ( t , t^{'} )
\int d x \; d x^{'} \; f ( x ) f ( x^{'} ) \nonumber \\
\{ \psi_L^\dagger ( x , t ) \psi_R ( x^{'} , t^{'} ) 
[ \; \langle \langle \psi_L ( x , t ) \psi_R^\dagger ( x^{'} , t^{'} ) 
\rangle \rangle 
\nonumber \\
+  \langle \langle
\psi_R ( x , t ) \psi_L^\dagger ( x^{'} , t^{'} ) \rangle \rangle 
e^{ 2 i q_F ( x + x^{'} )} \; ]
\nonumber \\
- \psi_L^\dagger ( x^{'} , t^{'} ) \psi_R ( x , t )  [ \; 
\langle \langle \psi_R^\dagger ( x , t ) \psi_L ( x^{'} , t^{'} ) \rangle
\rangle 
\nonumber \\
+ \langle \langle \psi_L^\dagger ( x , t ) \psi_R ( x^{'} , t^{'} )
\rangle \rangle  e^{ 2 i q_F ( x + x^{'} )} \; ] \}
\label{offidag4}
\end{eqnarray}
\noindent

In the local-time approximation, Eq.(\ref{offidag4})
takes the approximate form:

\begin{eqnarray} 
S_{LR}^{(d)} = \int d t \; \int d x \; d x^{'} \; f ( x ) \; f ( x^{'} )
\; \psi_L^\dagger ( x , t ) \psi_R ( x^{'} , t )
\nonumber \\
 e^{ i q_F ( x + x^{'} ) }
{\cal P} \biggl[ \frac{ \sin q_F ( x + x^{'} )}{ x + x^{'}  } \biggr]
\times 
\nonumber \\
\biggl[ 
G_q ( t + \frac{ x + x^{'}}{ v_F } , t ) + G_q 
(t,t - \frac{ x + x^{'}}{ v_F}) \biggr] \;\;\;\;\;\;
\label{offdiag5}
\end{eqnarray}
\noindent
Here the arguments of the two $G_q$'s to be added at 
the last line of Eq.(\ref{offdiag5}) are swapped  with respect to those
 appearing in  Eq.(\ref{loctb}). Hence, the sum now reads:
\begin{eqnarray} 
 G_q ( t + \frac{ x + x^{'}}{ v_F } , t ) + G_q ( t, t - \frac{ x + x^{'}}{ 
v_F}  )  
\label{offdiag6}  \\
  \approx 2
 \:\: \exp \biggl[  \frac{i}{2}
 \int_{t - (x+x^{'} )/v_F }^{t + (x+x^{'} )/v_F }\!\!\!  d \tau \: \dot{ \Phi} 
( \tau ) \biggr] \;\;  G_{q , A}^{\rm Ret} ( t^+ , t ) 
 \nonumber
\end{eqnarray}
This shows that the term of Eq.(\ref{offdiag5})   corresponds to 
the local potential $ V_\chi (t)$  in $H^{eff}_W $  (Eq.(\ref{pot})).  

The complex conjugate of Eq.(\ref{offdiag5}) is obtained, when working out
the contraction patterns giving raise to the $R-L$-backscattering term.

In conclusion, in constructing the effective potential arising from
capacitive coupling between arm and dot, we have identified two different
forward scattering terms, $ a)$ and  $ b)$ ,
 and two backscattering contributions, $ c)$ and  $ d)$. 
Terms $a) $ and $c )$ contribute to an electrostatic potential 
which is roughly  $\Phi$-independent and are  
disregarded in Section IV.
In Section IV, we just retain  terms  $b) $ and $d)$, and trade them 
for the local Hamiltonian  $ H^{eff}_W $ of Eq.(\ref{tham}).
Despite its apparent simplicity, $ H^{eff}_W $  is able to 
catch  the relevant asymptotic behavior of our physical system:
$\Phi$-dependent phase shifts and corresponding phase factors in the
reflection and transmission amplitudes. This is the key
motivation for using the simple local potential of   Eq.(\ref{pot}), rather
than the involute nonlocal scattering potential arising from the analysis
reported here.


\begin{thebibliography}{99}


\bibitem{divincenzo} D.P.DiVincenzo, Science {\bf 270}, 255 (1995).

\bibitem{nakamura} Y. Nakamura, Y. A. Pashkin and J. S. Tsai, Nature
{\bf 398}, 786 (1999).

\bibitem{flux} J. E. Mooij {\sl et al.}, Science {\bf 285}, 1036 (1999).

\bibitem{bayer} M. Bayer {\sl et al}, Science {\bf 291}, 451 (2001).

\bibitem{farhi} E.Farhi {\sl et al}, Science {\bf 292},
472 (2001); J. Pachos, P. Zanardi and M. Rasetti, Phys. Rev.{\bf A 61}, 
010305(R) (2000).

\bibitem{berry} M. V. Berry, Proc. R. Soc. London, Ser.A {\bf 392}, 45 (1984).

\bibitem{shenoy} G. Gaitan and S. R. Shenoy, Phys. Rev. Lett.{\bf 76}, 
4404 (1996).

\bibitem{fazio} G. Falci et al., Nature {\bf 407}, 355 (2000); Mahn-Soo Choi
quant-ph/0111019.

\bibitem{kastner} M. A. Kastner, Ann. Phys. 9, 885 (2000);
S. Sasaki {\sl et al.} Nature {\bf 405}, 764 (2000).

\bibitem{benoit} B. Jouault, G. Santoro and A. Tagliacozzo,
 Phys. Rev. {\bf B 61}, 10242 (2000);
L. P. Kouwenhoven {\sl et al.}, Science {\bf 278}, 1788 (1997).


\bibitem{loss}  J. Schliemann, D. Loss and A. H. MacDonald, Phys. Rev.
{\bf B 63}, 085311 (2001).

\bibitem{pazy} E. Pazy, I. D'Amico, P. Zanardi and F. Rossi, 
 Phys. Rev. {\bf B 64}, 195320 (2001).

\bibitem{beppe2} E. Ercolessi, G. Marmo, G. Morandi and N. Mukunda, Int.
Jou. Mod. Phys. {\bf A 16}, 5007 (2001).

\bibitem{ringarm} E. Buks, R. Shuster, M. Heiblum, D. Mahalu and V. Umansky,
Nature 391, 871 (1998); A. Yacoby, M. Heiblum, D. Mahalu and H. Shtrikman,
  Phys. Rev. Lett.{\bf 74}, 4047 (1995).


\bibitem{unpublished} B. Jouault  and A. Tagliacozzo,  unpublished.


\bibitem{referee} We are indebited to one of the Referees of our paper for 
having brought our attention to this point.

\bibitem{keldysh} L. V. Keldysh, Zh. Eksp. Teor. Fiz. {\bf 47}, 1515 (1964);
[Sov. Phys. JETP {\bf 20}, 1018 (1965)].


\bibitem{aronov} A. G. Aronov and Y. B. Lyanda-Geller, Phys. Rev. 
Lett.{\bf 70}, 343 (1993).

\bibitem{gaitan} F. Gaitan, J. Magn. Reson. {\bf 139} (1), 152 (1999).

\bibitem{aleiner} I.L.Aleiner, N.S.Wingreen and Y.Meir, Phys. Rev. Lett.
{\bf 79}, 3740 (1997).

\bibitem{pumping} J. E. Avron, A. Egart, G. M. Graf and L. Sadun, Phys. Rev.
Lett. {\bf 87}, 236601 (2001); Y. Levinson  {\sl et al.}, cond-mat/0010494.

\bibitem{avron0} J. E. Avron in ``Anomalies, Phases, Defects...'', M. Bregola,
G. Marmo, G. Morandi eds., Bibliopolis (Napoli) (1990).

\bibitem{rammer} J. Rammer and H. Smith, Rev. Mod. Phys. {\bf 58}, 323 (1986).


\end{thebibliography}
\end{document}